\documentclass[aps,prx,showpacs,floatfix,twocolumn,superscriptaddress,longbibliography,footinbib]{revtex4-1}

\usepackage{float}
\usepackage{xcolor}
\usepackage{bm}
\usepackage{todonotes}
\usepackage{verbatim}
\usepackage{soul}
\usepackage{braket}
\usepackage{amsfonts}
\usepackage{amsmath}
\usepackage{bm}
\usepackage{hyperref}
\hypersetup{
    colorlinks=true,
    linkcolor=blue,
    filecolor=magenta,      
    urlcolor=red,
    citecolor=blue,
}
\usepackage{glossaries}
\glsdisablehyper
\usepackage{sidecap}
\usepackage{verbatim}

\newacronym{QFI}{QFI}{Quantum Fisher Information}
\newacronym{nQFI}{nQFI}{normalized QFI}

\newacronym{RIXS}{RIXS}{resonant inelastic X-ray scattering}
\newacronym{XAS}{XAS}{X-ray absorption spectroscopy}
\newacronym{INS}{INS}{inelastic neutron scattering}
\newacronym{ARPES}{ARPES}{angle-resolved photoemission}
\newacronym{EELS}{EELS}{electron energy loss spectroscopy} 
\newacronym{STS}{STS}{scanning tunneling spectroscopy}
\newacronym{SOC}{SOC}{spin-orbit coupling}
\newacronym{QSL}{QSL}{quantum spin liquid}
\newacronym{XFEL}{XFEL}{X-ray free electron laser}
\newacronym{ED}{ED}{exact diagonalization}
\newacronym{EHM}{EHM}{extended Hubbard model}
\newacronym{EPC}{EPC}{electron-phonon coupling}
\newacronym{CDW}{CDW}{charge-density wave}
\newacronym{SDW}{SDW}{spin-density wave}
\newacronym{KH}{KH}{Kramers-Heisenberg}
\newacronym{QMC}{QMC}{quantum Monte Carlo}
\newacronym{UCL}{UCL}{ultra-short core hole lifetime}
\newacronym{DMRG}{DMRG}{density matrix renormalization group}
\newacronym{DMFT}{DMFT}{dynamical mean-field theory}
\newacronym{METTS}{METTS}{minimally entangled typical thermal states}
\newacronym{AFM}{AFM}{antiferromagnetic}
\newacronym{DCA}{DCA}{dynamical cluster approximation}
\newacronym{BCS}{BCS}{Bardeen, Cooper, and Schrieffer}
\newacronym{BZ}{BZ}{Brillouin zone}
\newacronym{1D}{1D}{one-dimensional}
\newacronym{2D}{2D}{two-dimensional}
\newacronym{3D}{3D}{three-dimensional}
\newacronym{trRIXS}{trRIXS}{time-resolved RIXS}
\newacronym{LCLS}{LCLS}{Linac Coherent Light Source}

\newcommand{\q}{\bm{q}}
\newcommand{\rv}{\bm{r}}

\usepackage{xr-hyper}
\begin{document}

\title{Observation of multipartite spin entanglement in a cuprate chain}

\author{S. F. R. TenHuisen}
\affiliation{Department of Physics, Harvard University, Cambridge, Massachusetts 02138, USA}
\affiliation{John A. Paulson School of Engineering and Applied Sciences, Harvard University, Cambridge, Massachusetts 02138, USA}

\author{Z. Shen}
\affiliation{Department of Chemistry, Emory University, Atlanta, GA 30322, USA}

\author{V. Bhartiya}
\affiliation{NSLS-II, Brookhaven National Laboratory, Upton, New York 11973, USA}

\author{V. Menon}
\affiliation{Department of Physics, Harvard University, Cambridge, Massachusetts 02138, USA}

\author{P. Sharma}
\affiliation{Department of Chemistry, Emory University, Atlanta, GA 30322, USA}

\author{H. Padma}
\affiliation{Department of Physics, Harvard University, Cambridge, Massachusetts 02138, USA}

\author{Z. Guan}
\affiliation{Department of Physics, Harvard University, Cambridge, Massachusetts 02138, USA}

\author{W. He}
\author{M. K. Lajer}
\affiliation{Condensed Matter Physics and Materials Science Department, Brookhaven National Laboratory, Upton, New York 11973, USA}

\author{J. Li}
\affiliation{NSLS-II, Brookhaven National Laboratory, Upton, New York 11973, USA}

\author{D. Banerjee}
\affiliation{Department of Physics and Astronomy, The University of Tennessee, Knoxville, Tennessee 37996, USA}
\affiliation{Institute for Advanced Materials and Manufacturing, University of Tennessee, Knoxville, Tennessee 37996, USA}

\author{J. Pelliciari}
\affiliation{NSLS-II, Brookhaven National Laboratory, Upton, New York 11973, USA}

\author{I. A. Zaliznyak}
\affiliation{Condensed Matter Physics and Materials Science Department, Brookhaven National Laboratory, Upton, New York 11973, USA}

\author{G. D. Gu}
\affiliation{Condensed Matter Physics and Materials Science Department, Brookhaven National Laboratory, Upton, New York 11973, USA}

\author{M. D. Lukin}
\affiliation{Department of Physics, Harvard University, Cambridge, Massachusetts 02138, USA}

\author{S. Johnston}
\affiliation{Department of Physics and Astronomy, The University of Tennessee, Knoxville, Tennessee 37996, USA}
\affiliation{Institute for Advanced Materials and Manufacturing, University of Tennessee, Knoxville, Tennessee 37996, USA}

\author{M. P. M. Dean}
\affiliation{Condensed Matter Physics and Materials Science Department, Brookhaven National Laboratory, Upton, New York 11973, USA}

\author{V. Bisogni}
\affiliation{NSLS-II, Brookhaven National Laboratory, Upton, New York 11973, USA}

\author{Y. Wang}\email[]{yao.wang@emory.edu}
\affiliation{Department of Chemistry, Emory University, Atlanta, GA 30322, USA}

\author{M. Mitrano}\email[]{mmitrano@g.harvard.edu}
\affiliation{Department of Physics, Harvard University, Cambridge, Massachusetts 02138, USA}

\date{\today}

\begin{abstract}
Quantum materials are believed to host highly entangled states of matter, but probing and quantifying such entanglement has long remained experimentally elusive. Here, we use \gls*{RIXS} as a probe of multipartite spin entanglement. By projecting the \gls*{RIXS} cross section onto the appropriate spin excitation channel, we show that the Quantum Fisher Information can be extracted directly from the measured spin fluctuation spectrum. We detect at least 7-partite spin entanglement in the model one-dimensional cuprate Sr$_2$CuO$_3$, revealing one of the largest entanglement depths yet reported in a solid. The observed entanglement depth agrees with Hubbard-model expectations and persists to elevated temperatures. Our results establish \gls*{RIXS} as an experimental probe of many-body spin entanglement and provide a platform for extensions to charge and orbital sectors and nonequilibrium settings.
\end{abstract} 

\maketitle

\section{Introduction}

Many-body entanglement is a fundamental ingredient in the physics of quantum materials~\cite{Amico2008entanglement,Keimer2017physics}. In frustrated magnets, strong quantum correlations suppress classical order and stabilize highly entangled states with long-range coherence and fractionalized excitations \cite{broholm2020quantum,Balents2010spin,Savary2016quantum,Wen2017colloquium}. Fractional quantum Hall states likewise arise from highly entangled composite-fermion wavefunctions \cite{Biddle2001entanglement,Zozulya2007bipartite,Li2008entanglement,Read2000paired}, while pseudogap and strange-metal behavior in unconventional superconductors have similarly been linked to entangled electronic states \cite{Bippus2025entanglement,Legros2019universal}. Quantum critical systems should exhibit entanglement over broad length and time scales, which amplifies quantum fluctuations and reshapes the response to external perturbations~\cite{Hertz1976quantum,Sachdev2011Quantum,Si2010heavy}.  Direct signatures of entanglement can reveal phases for which conventional symmetry-breaking order parameters are absent or ambiguous, as in quantum spin liquids \cite{broholm2020quantum,Takagi2019concept}, fluctuating superconductors \cite{Fradkin2015colloquium}, and many nonequilibrium states of matter \cite{delatorre2021nonthermal,Bloch2022strongly}. Quantitative knowledge of entanglement also informs whether a material can serve as a resource for quantum information systems
\cite{nielsen2010quantum,Degen2017quantum}. More broadly, entanglement metrics define the microscopic correlation patterns that quantum simulators must reproduce \cite{Zhao2025entanglement}, and thus the gate requirements needed to emulate real materials \cite{Semeghini2021probing,Bluvstein2022quantum,Manovitz2025quantum}. However, in contrast to isolated quantum systems used in quantum metrology and quantum information science, direct experimental probing of entanglement in solid-state quantum materials is a challenging task and an exciting frontier of modern physics~\cite{Vedral2008quantifying,Amico2008entanglement,Laflorencie2016quantum,Bouwmeester2000the,Fang2025amplified}.

\begin{figure*}
    \centering
    \includegraphics[width=0.8\linewidth]{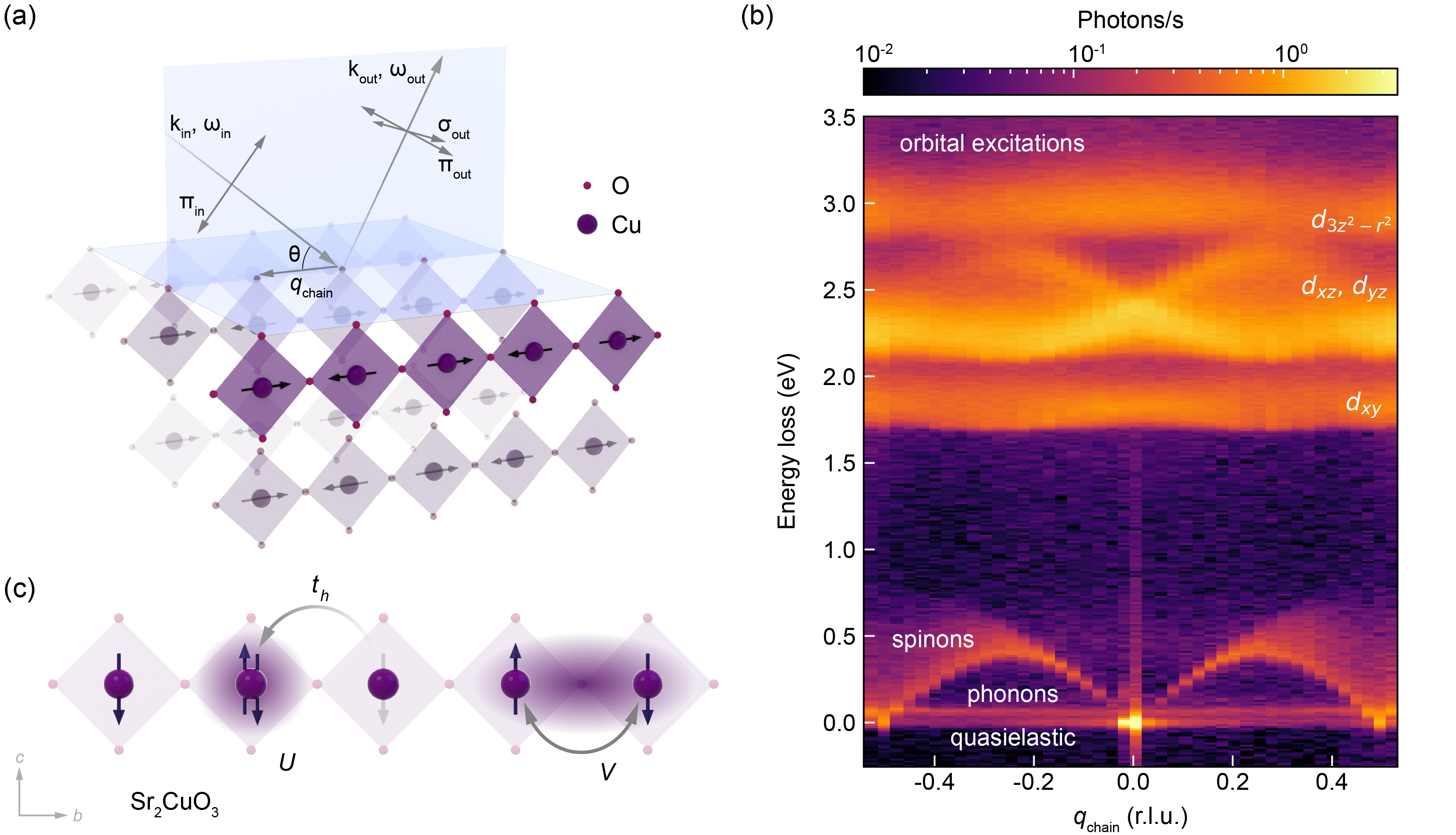}
    \caption{Collective excitations of the cuprate chain Sr$_2$CuO$_3$. (a) Sketch of the Cu $L_3$-edge \gls*{RIXS} experiment. We use incident $\pi$-polarized light to maximize the scattering cross section, and tune the momentum transfer along the chain direction by varying the sample angle, $\theta$. $\mathrm{k_{in}}$ ($\mathrm{k_{out}}$) and $\mathrm{\hbar\omega_{\rm in}}$ ($\mathrm{\hbar\omega_{out}}$) denote initial (final) momentum and energy of the X-ray photons. (b) Cu $L_3$-edge \gls*{RIXS} intensity map along the chain direction at 35~K, measured by tuning $\hbar\omega_{\rm in}$ to the maximum of the Cu $L_3$ X-ray absorption peak. Below 1~eV, the spectra exhibit dispersing two-spinon excitations; between 1.7 and 3.2~eV they reveal a manifold of $d$-orbital excitations. Owing to spin-orbital separation, $d_{xz}$ and $d_{yz}$ excitations become propagating orbitons and exhibit dispersive behavior. (c) Sr$_2$CuO$_3$ chains can be described as a 1D extended Hubbard model with hopping amplitude $t_h$, on-site Coulomb repulsion $U$, and intersite Coulomb interaction $V$.}
    \label{fig:RIXS_data}
\end{figure*}

A practical route to probe entangled states in solids relies on entanglement witnesses~\cite{Coffman2000distributed,amico2004dynamics,Roscilde2004studying,Brukner2006crucial,Amico2006divergence,Pezz2009,Hyllus2012,Toth2014}, operators whose expectation values are directly linked to experimental observables. Like Bell inequalities, these witnesses certify entanglement through the violation of bounds satisfied by all separable states, without requiring full state tomography. A particularly powerful example is the \gls*{QFI}, which certifies multipartite entanglement and can be extracted directly from the dynamical susceptibility~\cite{Hauke2016measuring,Hyllus2012}. Recent \gls*{INS} and muon spin rotation experiments have used the \gls*{QFI} to analyze quantum spin chains \cite{Mathew2020experimental,Scheie2021witnessing,Laurell2021quantifying}, proximate quantum spin liquids \cite{Pratt2022spin,Scheie2024proximate,Wu2025spin}, and heavy-fermion systems \cite{Mazza2026quantum}, establishing a concrete protocol for quantifying the depth and distribution of spin entanglement in quantum magnets~\cite{Laurell2025witness}. These developments call for entanglement witnesses in spectroscopic probes that can access a broader range of microscopic degrees of freedom across a wider class of quantum materials \cite{shen2025witnessing,ren2024witnessing,liu2025entanglement}.

\Gls*{RIXS} is a natural platform for probing entangled states in quantum materials. Over the past two decades, it has become an effective probe of charge, orbital, and spin degrees of freedom, with unique advantages for small samples, thin films, and \gls*{2D} materials~\cite{Ament2011resonant,Mitrano2024exploring,deGroot2024resonant}. Its scattering amplitude is phase coherent and highly sensitive to the presence of distinct electronic pathways \cite{Schulke2007electron}, as underscored by pioneering RIXS-interferometry studies of dimer and trimer iridates~\cite{Revelli2019resonant,Revelli2022quasimolecular,Magnaterra2023RIXS,Magnaterra2025quasimolecular}. Turning \gls*{RIXS} into an entanglement probe, however, is not straightforward, particularly compared with \gls*{INS}. Because \gls*{RIXS} couples to multiple microscopic degrees of freedom, its excitation operator is more complex than a local spin flip, and the corresponding \gls*{QFI} bounds become material- and geometry-dependent. Moreover, \gls*{RIXS} does not generally yield an absolute scattering cross section, so its use as an entanglement witness demands careful normalization. Local spin degrees of freedom provide a natural starting point for witnessing entanglement with \gls*{RIXS}. Yet, despite the widespread use of RIXS as a probe of magnetic excitations, spin entanglement has not yet been detected with this technique.

Here, we establish \gls*{RIXS} as a probe of spin entanglement in strongly correlated electron systems. We focus on copper oxides, where two decades of scattering experiments \cite{Braicovich2010magnetic,LeTacon2011Intense,Dean2012spin} and a well-characterized cross section \cite{Ament2009theoretical,Haverkort2010theory,Robarts2021dynamical} provide stringent experimental and theoretical benchmarks. Using high-resolution \gls*{RIXS} and theoretical calculations on the prototypical \gls*{1D} cuprate Sr$_2$CuO$_3$, we project the measured signal onto the spin dynamical structure factor (accounting for multiple spin channels and core-hole effects) and extract the \gls*{QFI} through a sum-rule integral. The resulting momentum-resolved \gls*{QFI} reveals multipartite spin entanglement, at least 7-partite at low temperature, in quantitative agreement with theory. Our results establish \gls*{RIXS} as a spectroscopic probe of many-body entanglement in quantum materials and open a path to detecting entanglement in charge and orbital sectors, as well as in nonequilibrium states of matter.

\section{A paradigmatic 1D cuprate}\label{sec:paradigmatic}

Sr$_2$CuO$_3$ is an ideal platform for probing many-body entanglement with \gls*{RIXS} [Fig.~\ref{fig:RIXS_data}(a)]. This half-filled Mott insulator consists of chains of corner-sharing CuO$_4$ plaquettes aligned along the $b$ axis \cite{Motoyama1996magnetic,Kojima1997reduction}, giving rise to a local structure, chemical environment, and microscopic interactions closely related to those of the superconducting cuprates. At the same time, the interchain coupling is extremely weak, rendering the chains effectively decoupled and realizing nearly ideal \gls*{1D} physics \cite{Ami1995magnetic,Maiti1998electronic,Walters2009effect,Sergeicheva2017unusual}.

In one dimension, strong quantum fluctuations suppress long-range order and stabilize a critical, strongly entangled ground state, while the elementary excitations fractionalize through spin--charge and spin--orbital separation \cite{Walters2009effect,schlappa2012spin,Neudert1998manifestation}. These features are directly visible in our high-resolution ($\sim35$~meV) Cu $L_3$-edge \gls*{RIXS} spectra along the chain direction [Fig.~\ref{fig:RIXS_data}(b)]. Between $1.5$ and $3.0$~eV, we observe dispersive $d_{xz}$ and $d_{yz}$ orbitons, consistent with prior results in Ref.~\onlinecite{schlappa2012spin}, providing a clear signature of spin--orbital separation. Below $\sim0.8$~eV, the spectra show a prominent continuum of strongly dispersing magnetic excitations, corresponding to pairs of spin-$1/2$ quasiparticles propagating along the chain \cite{Giamarchi2004quantum}. This two-spinon continuum is consistent with prior work identifying Sr$_2$CuO$_3$ as a near-ideal spin-$1/2$ chain with a large superexchange scale, $J \simeq 250$~meV \cite{schlappa2012spin,Schlappa2018probing}. At still lower energies, we resolve bond-stretching phonons, as observed across other cuprate families. Related corner-sharing quasi-\gls*{1D} cuprates show the same underlying phenomenology. Ca$_2$CuO$_3$ displays dispersive two-spinon and orbiton excitations in the undoped chain limit, while doped Ba$_2$CuO$_{3+\delta}$ reveals how the two-spinon spectrum evolves away from half filling \cite{Fumagalli2020mobile,Li2025doping}. Further experimental details are provided in Appendices~\ref{ap:ex_growth} and \ref{ap:ex_methods}, and Supplemental Material Secs.~S1 and S2 \cite{Supplementary}.

The reduced dimensionality of Sr$_2$CuO$_3$ makes it particularly well suited for quantitative comparison with theory, which is far more controlled in one dimension than in higher dimensions. Its low-energy physics and the corresponding \gls*{RIXS} intensity are captured by the \gls*{EHM} with nearest-neighbor hopping $t_h \simeq 560~\mathrm{meV}$, on-site interaction $U = 8t_h$, and attractive nearest-neighbor interaction $V = -t_h$ [Fig.~\ref{fig:RIXS_data}(c) and Appendix~\ref{app:numerical_methods}], consistent with earlier work \cite{Neudert1998manifestation,Neudert1999electronic,Neudert2000four} and recent photoemission and RIXS measurements \cite{chen2021anomalously,Li2025doping}. Here, $V<0$ denotes an effective attractive Coulomb interaction at low energy, distinct from the bare repulsive intersite Coulomb interaction. All model parameters used are fixed by independent spectroscopic experiments. In the large-$U$ limit, the magnetic sector reduces to a Heisenberg model~\cite{Nocera2016magnetic} with $J \sim 250$~meV (Supplemental Material Sec.~S5~\cite{Supplementary}). The orbital sector is not captured within this effective description and would require a multiband treatment beyond the scope of this work \cite{Wohlfeld2013microscopic}.

\section{Witnessing multipartite spin entanglement}\label{sec:QFI_defns}

The main focus of this work is multipartite spin entanglement. Because entanglement is a property of the wavefunction, we characterize it through the notion of $m$-producibility. A pure spin state $\ket{\Psi_{m\text{-}\rm{prod}}}$ is $m$-producible if it factorizes into non-overlapping partitions,
\begin{equation}\label{eq:spinState}
\ket{\Psi_{m\text{-}\rm{prod}}} = \ket{\Phi_1}\otimes \ket{\Phi_2}\otimes\cdots\otimes \ket{\Phi_l},
\end{equation}
where each $\ket{\Phi_i}$ is an irreducible many-body state defined on a partition $P_i$ with $|P_i| \le m$. The entanglement depth is then $m$ if the state is $m$-producible but not $(m-1)$-producible, meaning that it cannot be decomposed into product wavefunctions containing fewer than $m$ spins. This definition of entanglement in a pure state is generalizable to mixed quantum states (Supplemental Material Sec.~S6~\cite{Supplementary}), such as the finite-temperature spin chain we consider here. Larger entanglement depth therefore reflects entanglement distributed across increasingly large portions of the system and provides a direct measure of many-body complexity.

We use the \gls*{QFI} density as a witness of multipartite entanglement. For a mixed state of an ensemble of $N$ particles, the \gls*{QFI} density is defined as
\begin{equation}
f_Q(\rho,\hat O)=\frac{2}{N}\sum_{n,n'}\frac{(p_n-p_{n'})^2}{p_n+p_{n'}}|\bra{n}\hat O\ket{n'}|^2,\label{eq:QFI_def_mixed}
\end{equation}
where $\rho=\sum_n p_n\ket{n}\bra{n}$ is the density matrix and $\hat O=\sum_{i=1}^N \hat O_i$ a sum of local Hermitian operators with bounded spectra~\cite{Escher2011general,Braunstein1994statistical}. For an $m$-producible spin state, the \gls*{QFI} density is bounded by \cite{Hyllus2012}
\begin{equation}
f_Q \le m (h_{\text{max}}-h_{\text{min}})^2,
\end{equation}
where $h_{\text{max}}$ and $h_{\text{min}}$ are the extremal eigenvalues of the local observable $\hat O_i$. For spins with local moment $S = 1/2$, these become $h_{\text{max}}=1/2$ and $h_{\text{min}}=-1/2$, and a value exceeding $m$ therefore certifies at least $(m+1)$-partite entanglement. The \gls*{QFI} thus provides a Bell-like test for the presence of multipartite entanglement with a given depth.

Crucially, the \gls*{QFI} density is experimentally accessible through a sum rule over the dynamical response \cite{Hauke2016measuring},
\begin{equation}\label{eq:QFIIntegral}
f_Q(q,T) = 4 \int_{-\infty}^{\infty} \tanh^2\!\bigg(\frac{\hbar \omega}{2 k_B T}\bigg)\, S(q,\hbar\omega,T)\, d(\hbar\omega),
\end{equation}
where $\tanh(\hbar\omega/2k_BT)$ acts as a filter that suppresses thermal contributions and precisely isolates quantum fluctuations. $S(q,\hbar\omega,T)$ is the momentum-, frequency\mbox{-,} and temperature-dependent dynamical structure factor~\cite{Hauke2016measuring}. For spin fluctuations in an isotropic magnet,
\begin{equation}
S(q,\hbar\omega,T) = \frac{1}{2\pi\hbar}\int dt\, e^{-i\omega t} \sum_i e^{i \mathbf{q} \cdot \mathbf{r}_i}\langle S_0^{\alpha}(0) S_i^{\alpha}(t) \rangle,
\end{equation}
where $S_i^\alpha$ denotes the $\alpha$ component of the spin operator ($\alpha=x,y,z$) on site $i$ at position $\mathbf{r}_i$ and time $t$, and the brackets denote a thermal ensemble average. Total moment sum rules require that $\int dq \int d(\hbar\omega) S(q,\hbar\omega) = S(S+1)/3$ for each of the three components of $S(q,\hbar\omega)$~\cite{Lorenzana2005sum}. For simplicity, we omit the explicit temperature dependence of $S(q,\hbar\omega)$ from now on. Further details of the connection between $f_Q(q,T)$ and the dynamical spin structure factor $S(q,\hbar\omega)$ are given in Supplemental Material Sec. S6~\cite{Supplementary}.

The correlation function $S(q,\hbar\omega)$ can be extracted from INS through well-established data treatment procedures, including phonon subtraction, normalization through the atomic form factor, appropriate scaling with $q$, and often the use of overall sum rules \cite{Lorenzana2005sum}. In \gls*{RIXS}, by contrast, the richer scattering cross section contains multiple excitation channels and correlation functions and therefore does not map directly onto $S(q,\hbar\omega)$. Extracting the dynamical spin structure factor thus requires an additional, theory-assisted cross-section projection, which in the present work is obtained from an experimentally benchmarked model of Sr$_2$CuO$_3$ (as discussed in the next section). For SU(2)-symmetric systems such as the isotropic spin-$1/2$ chain studied here, one can conceptualize direct \gls*{RIXS} in terms of spin creation and destruction operators $\langle S_0^{-}(0) S_i^{+}(t) \rangle$, corresponding to $\Delta S=1$ excitations \cite{Ament2009theoretical,Haverkort2010theory,jia2016using}. By contrast, indirect \gls*{RIXS} processes can probe higher-order spin-conserving $\Delta S=0$ correlations \cite{Forte2011doping,jia2016using,Klauser2011}, which do not enter the present \gls*{QFI} formulation. Witnessing spin entanglement with \gls*{RIXS} therefore requires isolating the $\Delta S=1$ spin correlations from the scattered signal, a step at which X-ray and neutron protocols diverge because of their distinct cross sections.

\section{RIXS intensity analysis}\label{sec:projection}

\begin{figure*}
    \centering
    \includegraphics[width=0.95\linewidth]{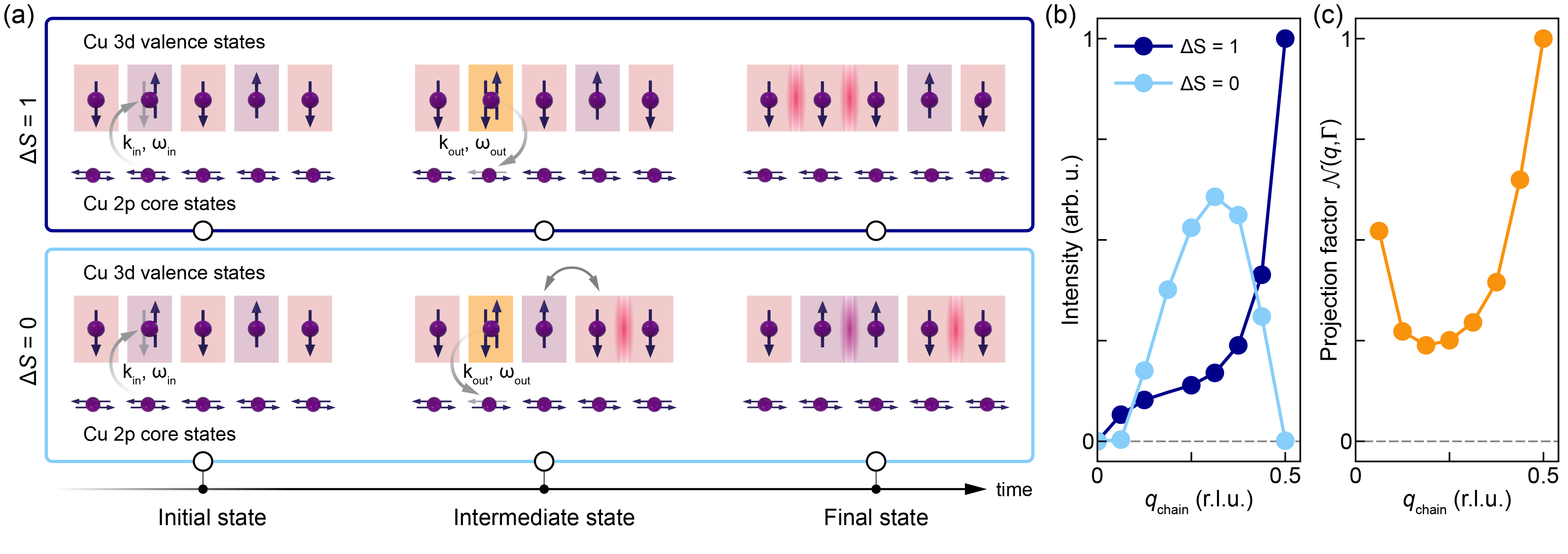}
    \caption{Two distinct two-spinon excitation pathways. (a) Following resonant X-ray absorption, two-spinon excitations are generated via spin-flip ($\Delta S = 1$) and spin-conserving ($\Delta S = 0$) channels. The former arise from a direct spin flip upon de-excitation from the intermediate state, while the latter involve a spin rearrangement during the intermediate state. (b) Calculated \gls*{RIXS} intensities of $\Delta S = 1$ (dark blue) and $\Delta S = 0$ (light blue) channels in Sr$_2$CuO$_3$ as a function of momentum transfer along the chain direction. (c) Calculated projection factor $\mathcal{N}(q,\Gamma)$ relating the total \gls*{RIXS} response to $S(q,\hbar\omega)$ for $\Gamma=0.5t_h$.}

    \label{fig:RIXS_corrections}
\end{figure*}

As spins do not couple directly to light, \gls*{RIXS} probes spin excitations through two distinct pathways mediated by the intermediate state \cite{Ament2009theoretical,Ament2011resonant,Haverkort2010theory}. At the Cu $L_3$ absorption edge, two-spinon excitations arise through both spin-flip (direct \gls*{RIXS}) and spin-conserving (indirect \gls*{RIXS}) channels [Fig.~\ref{fig:RIXS_corrections}(a)]. A core electron is resonantly promoted into the valence band, creating a transient doubly occupied site (orange). In the direct channel (dark-blue box), a valence electron refills the core hole and, through spin--orbit coupling in the intermediate state, leaves behind a flipped spin, generating spin-flip $\Delta S=1$ excitations (2 spinons with the same spin alignment). In the indirect channel (light-blue box), the spin system evolves under a locally modified superexchange Hamiltonian due to the transient $3d^{10}$ configuration. During this intermediate state lifetime, spins may scatter from the excited site, leaving behind spin-conserving $\Delta S=0$ excitations (2 spinons with opposite spin alignment) after a valence electron refills the core hole without flipping a spin \cite{Bisogni2014femtosecond}.

To extract the momentum-dependent structure factor, we must separate the two spin scattering channels. In $d^{9}$ systems, including many cuprates, spin-flip and spin-conserving contributions can be distinguished experimentally in certain scattering geometries through polarization selection rules, with spin-conserving scattering occurring only in the polarization-conserving channel and spin-flip scattering occurring only in the polarization-rotating channel (Supplemental Material Fig.~S4~\cite{Supplementary}). In Sr$_2$CuO$_3$, however, the scattering geometry imposed by the natural cleavage plane prevents this approach: $\sigma$-polarized light is perpendicular to the valence $d_{x^2-y^2}$ orbital at all angles, causing both the absorption and emission cross sections for $\sigma$-polarized X-rays to vanish (Supplemental Material Sec.~S3A \cite{Supplementary}), while spin-conserving and spin-flip scattering processes both contribute to the cross section for $\pi$-polarized incident and emitted X-rays. Future work can take advantage of the full tensorial nature of the \gls*{RIXS} cross section to experimentally isolate the spin-conserving and spin-flip contributions, as each \gls*{RIXS} spin channel has a distinct symmetry for its azimuthal intensity dependence. This has been demonstrated for magnons and bimagnons in 2D cuprates \cite{Kang2019Resolving} and can be generalized to other materials platforms. 

Here, we determine the relative spin-flip and spin-conserving cross sections for Sr$_2$CuO$_3$ by exploiting the fact that Sr$_2$CuO$_3$ is well described by the \gls*{EHM} and using \gls*{ED} to compute the momentum-dependent \gls*{RIXS} intensity $\mathcal{I}_{\text{RIXS}}(q,\hbar\omega_{\rm in},\hbar\omega)$ for each spin channel [Appendix Fig.~\ref{fig:theory_RIXS_Sqw}]. The channel-resolved intensity depends on the intermediate-state core-hole lifetime and can be written as a fundamental scattering amplitude multiplied by a polarization- and geometry-dependent matrix element (Appendix~C; Supplemental Material Sec.~S3 \cite{Supplementary}). Although the two channels carry comparable total integrated intensity, their momentum dependencies are markedly different [Fig.~\ref{fig:RIXS_corrections}(b)]~\cite{Bisogni2014femtosecond}. At $|q_{\mathrm{chain}}|=0.5$~r.l.u., the spectral weight arises entirely from $\Delta S=1$ two-spinon excitations, whereas near $|q_{\mathrm{chain}}|\simeq 0.25$~r.l.u.\ it is dominated by $\Delta S=0$ excitations. Thus, approaching the antiferromagnetic wavevector progressively isolates the $\Delta S=1$ channel relevant to the spin structure factor.

In the ultrashort core-hole lifetime limit, the direct spin-flip ($\Delta S=1$) \gls*{RIXS} channel approaches $S(q,\hbar\omega)$~\cite{Ament2009theoretical,jia2016using}. In practice, the finite lifetime of the localized core hole in the intermediate state redistributes spectral weight among spin excitations at different momenta and energies [see Appendix Fig.~\ref{fig:corehole_lifetime_correction}]. Therefore, we further project the $\Delta S=1$ \gls*{RIXS} channel onto $S(q,\hbar\omega)$ assuming an inverse core-hole lifetime of $\Gamma \sim 0.5t_h$ ($t_h$ being the hopping amplitude), as estimated from the Cu $L_3$-edge \gls{XAS} (Supplemental Material Sec.~S1 \cite{Supplementary}). 

The full projection factor for Sr$_2$CuO$_3$, which accounts for both magnetic excitation channels and the finite core-hole lifetime, is shown in Fig.~\ref{fig:RIXS_corrections}(c). We define this projection factor as
\begin{equation}\label{eq:norm}
    \mathcal{N}(q, \Gamma) = \frac{\int_{-\infty}^{\infty} S(q,\hbar\omega)\, d(\hbar\omega)}{\int_{-\infty}^{\infty} \mathcal{I}_{\text{RIXS}}(q,\hbar\omega_{\rm in}, \hbar\omega)\, d(\hbar\omega) }, 
\end{equation}
where $\mathcal{I}_{\text{RIXS}}(q,\hbar\omega_{\rm in},\hbar\omega)$ is the total \gls*{RIXS} intensity from small-cluster \gls*{ED} calculations and $S(q,\hbar\omega)$ is the spin structure factor of an extended Hubbard chain calculated with \gls*{DMRG} (see Appendix ~\ref{app:numerical_methods} for details). In evaluating $\mathcal{N}(q, \Gamma)$, we fix $\hbar\omega_{\rm in}$ to the energy of the Cu $L_3$-edge resonance. This factor converts the sum-rule integral of the measured \gls*{RIXS} intensity at each momentum into the corresponding sum-rule integral of $S(q,\hbar\omega)$ required to extract the \gls*{QFI}. Finally, to place the spin \gls*{QFI} on an absolute scale, we normalize the projected RIXS intensity by fixing the total inelastic spin spectral weight integrated over energy loss and the entire Brillouin zone to $S(S+1)/3$, as required by the total-moment sum rule \cite{Lorenzana2005sum, Scheie2021witnessing,Scheie2023reconstructing,Laurell2021quantifying,Boothroyd2020,Shirane_Shapiro_Tranquada_2002}.

The polarization- and geometry-dependent matrix elements largely set the relative weights of spin-flip and spin-conserving excitations in \gls*{RIXS}, and can be well-approximated for any material and scattering geometry within an atomic-model framework (Supplemental Material Sec.~S3A \cite{Supplementary}). The core-hole lifetime, in turn, is fixed by the chosen absorption edge (Appendix~C; Supplemental Material Sec.~S1 \cite{Supplementary}). Thus, while the projection factor in this work is benchmarked using a specific Hamiltonian, its inputs are largely spectroscopic and geometric, making the procedure extendable across cuprates at the Cu $L_3$ edge and adaptable to other absorption edges, provided that the projection factor is recalculated or experimentally constrained for each material and scattering geometry (Supplemental Material Secs.~S7 and S8~\cite{Supplementary}). Future developments in absolute-intensity calibration against well-characterized scattering standards \cite{Muller2006absolute} or suitable Bragg reflections (such as the (0,0,2) Bragg reflection of Bi$_2$Sr$_2$CaCu$_2$O$_{8+\delta}$) \cite{Abbamonte2005spatially} could yield an experimentally determined absolute RIXS cross section and further reduce reliance on material-specific theoretical modeling.

\section{Multipartite spin entanglement in \texorpdfstring{$\mathrm{Sr_2CuO_3}$}{Sr2CuO3}}\label{sec:results}

\begin{figure*}
    \centering
    \includegraphics[width=0.95\linewidth]{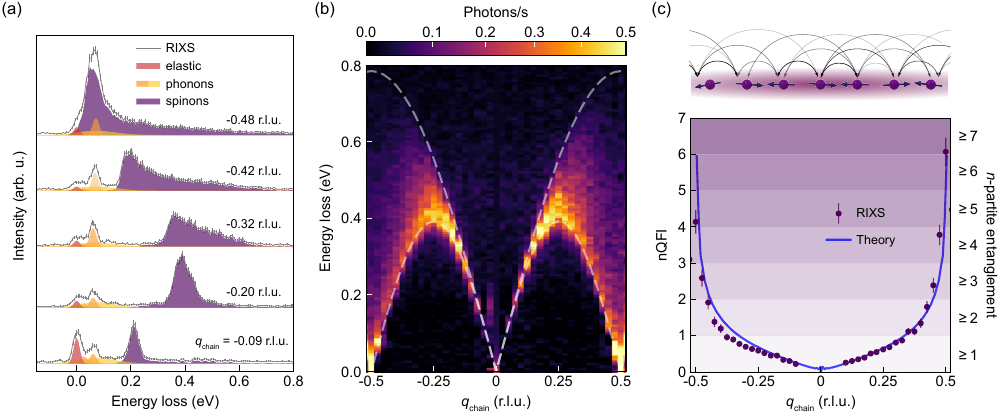}
    \caption{Observation of 7-partite spin entanglement. (a) Experimental RIXS spectra at 35~K at selected momenta. Shaded areas correspond to the quasielastic, phonon, multiphonon, and two-spinon continuum contributions. The energy resolution is $\sim$35~meV. (b) Experimental \gls*{RIXS} intensity map as a function of momentum and energy loss, after subtracting elastic and phonon contributions and normalizing as described in the text. White dashed lines are the boundaries of the two-spinon continuum. (c) Normalized \gls*{QFI} (nQFI) extracted from the \gls*{RIXS} spectrum (dark purple), compared with theoretical expectations for an extended Hubbard chain of length $L=128$ (blue). Error bars are obtained by propagating the Poisson uncertainties on the integrated measured \gls*{RIXS} photon numbers. The normalized \gls*{QFI} exceeds 6 at $|q_{\mathrm{chain}}|=0.5$~r.l.u., witnessing at least 7-partite spin entanglement.}
    \label{fig:QFI_RIXS}
\end{figure*}

We now turn to the main result of this work, namely the observation of multipartite spin entanglement in Sr$_2$CuO$_3$. Below 150 meV, the \gls*{RIXS} spectra contain nondispersive contributions from the elastic line, phonons and their harmonics, together with a clear dispersing two-spinon continuum [Fig.~\ref{fig:QFI_RIXS}(a)]. To isolate the spin response, we first correct for self-absorption \cite{Wang2020} and normalize each spectrum to the expected angle-dependent intensity of the nondispersing $d_{3z^2-r^2}$ orbital, calculated within a single-site cluster, thereby accounting for variations of beam footprint and collection efficiency with sample angle (Supplemental Material Sec.~S2 \cite{Supplementary}). We use the $d_{3z^2-r^2}$ orbital as a standard candle because it remains localized, whereas the other orbital excitations disperse through spin--orbital separation and are not reliably captured by few-site cluster calculations \cite{Wang2019,Wohlfeld2013microscopic}. After fitting and subtracting the elastic and phonon contributions, the remaining signal is the two-spinon continuum [Fig.~\ref{fig:QFI_RIXS}(b)], with spectral weight concentrated near its lower edge around $|q_{\mathrm{chain}}|=0.25$ and $0.5$ r.l.u. (the continuum bounds for $J=250$ meV are overlaid in white). Weak intensity near $q_{\mathrm{chain}}=0$ at $\sim0.5$~eV, above the upper edge of the two-spinon continuum, is assigned to four-spinon excitations \cite{Caux2006four,Schlappa2018probing} and excluded from the \gls*{QFI} integral in Eq.~\eqref{eq:QFIIntegral}.

We then evaluate the normalized \gls*{QFI} (nQFI) integral at each momentum by weighting the measured \gls*{RIXS} intensity with the projection factors in Fig.~\ref{fig:RIXS_corrections}(c). The resulting momentum-dependent \gls*{QFI}, shown in Fig.~\ref{fig:QFI_RIXS}(c), peaks sharply at the zone boundary, $|q_{\mathrm{chain}}|=0.5$~r.l.u.\ ($|q_{\mathrm{chain}}|=\pi$), where it exceeds six and certifies an entanglement depth of at least 7 spins at 35~K. The large entanglement depth follows from the critical nature of the spin-$\tfrac{1}{2}$ chain, as its gapless ground state sustains quantum correlations over all length scales. The same criticality also implies that the \gls*{QFI} at the antiferromagnetic wavevector diverges logarithmically upon cooling~\cite{Menon2023}. Because the superexchange in Sr$_2$CuO$_3$ is exceptionally large, the chain remains effectively close to its zero-temperature critical ground state even at 35~K ($k_B T/J=0.013$), yielding a correspondingly large \gls*{QFI}.

The zone-boundary maximum of the QFI density arises due to the presence of staggered antiferromagnetic correlations. The QFI density is related to the coherent part of the fluctuations of operators with the general form $\hat{O}_{\q}=\sum_j^N \cos(\q\cdot\rv_j + \phi)S_j^{\alpha}$. Due to the antiferromagnetic nature of the interactions in Sr$_2$CuO$_3$, at $q=Q_\text{AF}$ the dominant staggered contributions add constructively and provide the most sensitive witness of entanglement. Away from $Q_\text{AF}$, the residual phase winds with distance, causing longer-range contributions to interfere destructively and yielding a progressively looser bound on the entanglement \cite{Simeth2026quantum}.

We emphasize that the observed entanglement depth is robust against modest uncertainties in the projection factor. At $|q_{\mathrm{chain}}|=0.5$~r.l.u., the magnetic \gls*{RIXS} spectral weight is overwhelmingly dominated by the $\Delta S=1$ channel and the projection factor is maximized. Although an ideal \gls*{1D} spin chain at zero temperature has divergent entanglement depth near $|q_{\mathrm{chain}}|=0.5$~r.l.u., finite momentum resolution necessarily averages over the vicinity of the zone boundary and thus yields a finite measured depth. After accounting for experimental resolution broadening, the extracted normalized \gls*{QFI} quantitatively follows the \gls*{EHM} result at every measured momentum, despite the simplicity of the single-band description.

The 7-partite entanglement witnessed in Sr$_2$CuO$_3$ is among the deepest reported in solids to date. Inelastic neutron scattering has witnessed at least four-partite entanglement in the Heisenberg chain KCuF$_3$ ($J=33.5$~meV, $k_B T/J=0.015$)~\cite{Scheie2021witnessing}, and tripartite entanglement in the XXZ chain Cs$_2$CoCl$_4$ ($J=0.23$~meV, $k_B T/J=0.026$)~\cite{Laurell2021quantifying}. Four-partite entanglement has also been reported in the proximate spin-liquid KYbSe$_2$~\cite{Scheie2024proximate}, while comparable depths have been reached in quantum-critical heavy-fermion metals, where high-resolution neutron measurements at millikelvin temperatures find $f_Q=8.2\pm0.9$ near a Kondo-destruction critical point~\cite{Mazza2026quantum}. By contrast, in two-dimensional cuprates, sum-rule integrations of uncalibrated inelastic neutron scattering data yield at most bipartite entanglement in the pseudogap regime~\cite{Bippus2025entanglement}, consistent with weaker quantum fluctuations and with superconductivity cutting off the low-temperature growth of spin entanglement. The comparison with two-dimensional cuprates shows that, even for comparable exchange scales, one-dimensional criticality is essential for enhancing the entanglement depth of the spin sector.

\begin{figure*}
    \centering
    \includegraphics[width=0.9\linewidth]{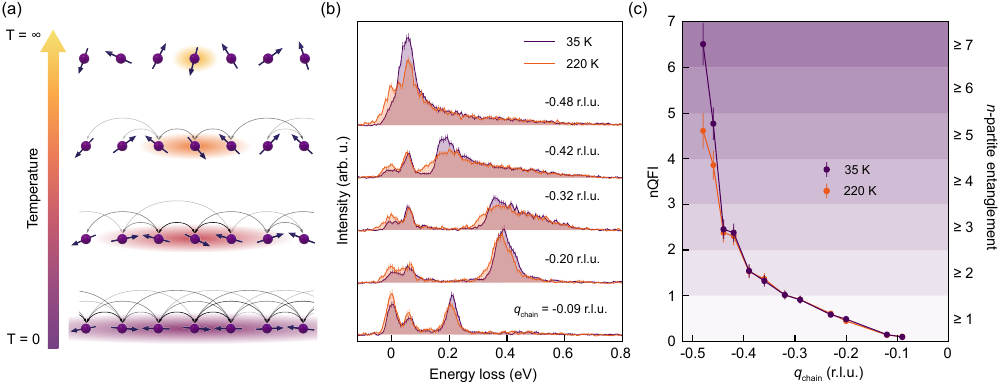}
    \caption{Multipartite spin entanglement at high temperature. (a) Sketch of the multipartite spin entanglement as a function of temperature. While at low temperature spins are highly entangled, upon increasing temperature quantum correlations become weaker and shorter range. (b) Low-energy \gls*{RIXS} spectra at selected momenta at 35~K (purple) and 220~K (orange). Spectra are shifted for clarity. With increasing temperature, the spectra are modestly reshaped. (c) Normalized \gls*{QFI} (nQFI) extracted from the \gls*{RIXS} data at 35~K (purple circles) and 220~K (orange circles). Error bars are obtained by propagating the Poisson uncertainties on the integrated measured \gls*{RIXS} photon numbers. The normalized \gls*{QFI} is slightly reduced at the zone boundary and remains above 4, witnessing at least 5-partite entanglement at 220~K.}
    \label{fig:QFI_Tdep}
\end{figure*}

\par We next examine the temperature dependence of the observed spin entanglement. A spin-$1/2$ Heisenberg chain is critical at $T=0$, but its entanglement is expected to weaken with increasing temperature, following a characteristic logarithmic scaling~\cite{Hauke2016measuring,Menon2023}~[Fig.~\ref{fig:QFI_Tdep}(a)]. We compare \gls*{RIXS} spectra measured at 35~K and 220~K [Fig.~\ref{fig:QFI_Tdep}(b)]. At 220~K, the two-spinon continuum broadens and loses spectral weight, particularly near the zone boundary $|q_{\mathrm{chain}}|=0.5$~r.l.u., where staggered correlations are strongest. Spectra at intermediate temperatures near $|q_{\mathrm{chain}}|=0.45$~r.l.u.\ evolve continuously (Supplemental Material Sec.~S4 \cite{Supplementary}), consistent with the absence of a phase transition. The corresponding normalized \gls*{QFI} is shown in Fig.~\ref{fig:QFI_Tdep}(c). Near the zone center, changes in the inferred entanglement depth are marginal and reflect the reduced sensitivity of the \gls*{QFI} away from the antiferromagnetic wavevector. Near the zone boundary, by contrast, the loss and broadening of two-spinon spectral weight suppress the \gls*{QFI} to a value still above 4, certifying at least 5-partite entanglement at 220~K. Sr$_2$CuO$_3$ thus retains substantial multipartite spin entanglement at elevated temperature, as its large $J$ keeps the chain in the quantum regime even at 220~K ($k_BT/J\simeq 0.08$, Supplemental Material Sec.~S9 \cite{Supplementary}).

\section{Conclusions and outlook}

Detecting multipartite spin entanglement with RIXS in Sr$_2$CuO$_3$ has broad implications for the microscopic understanding of cuprates and other quantum materials. Owing to their large magnetic energy scales, cuprates can sustain substantial spin entanglement up to elevated temperatures, plausibly beyond the \gls*{1D} limit. Extending entanglement witnesses to \gls*{2D} cuprates would place direct constraints on the interactions shaping their magnetic continua and shed light on the reported fractionalization of magnons into two spinons \cite{Martinelli2022fractional,Headings2010anomalous}, akin to observations in other \gls*{2D} square-lattice quantum antiferromagnets \cite{DallaPiazza2015fractional}. Because candidate resonating-valence-bond \cite{Anderson1987resonating,DallaPiazza2015fractional}, AF* \cite{Ho2001nature}, valence-bond-solid \cite{Sandvik2007evidence,Tang2013confinement,Shao2017nearly}, $\pi$-flux spin-liquid \cite{Bonetti2024quantum}, and disordered phases \cite{Sabharwal2025witnessing,shimokawa2025experimentally,sabharwal2026characterizing} differ in their entanglement structure, the \gls*{QFI} provides a direct route to distinguish them through their spectroscopic signatures. Combined with accurate theory, the momentum and doping dependence of the paramagnon \gls*{QFI} above $T_c$ could further constrain the putative quantum critical point \cite{Badoux2016change,Michon2019thermodynamic,Sachdev2003colloquium} and test whether the pseudogap and strange-metal regimes are highly entangled states of matter, as suggested by a recent analysis of inelastic neutron scattering data from Hg-based cuprates and YBa$_2$Cu$_3$O$_{6+x}$ \cite{Bippus2025entanglement}.

The present implementation extracts the spin \gls*{QFI} through a material- and geometry-specific mapping of the \gls*{RIXS} cross section onto $S(q,\hbar\omega)$, constrained by the benchmarked Hamiltonian of Sr$_2$CuO$_3$. A natural next step is to make this mapping increasingly reliant on purely experimental quantities. Entanglement witnesses may be formulated directly in terms of the full \gls*{RIXS} operator \cite{ren2024witnessing}, while momentum- and polarization-resolved measurements could yield higher-order correlators, including the two-particle cumulant reduced density matrix \cite{liu2025entanglement}. In parallel, absolute-intensity calibration against well-characterized scattering standards \cite{Muller2006absolute} or suitable Bragg reflections \cite{Abbamonte2005spatially} could place \gls*{RIXS} on an experimentally determined absolute scale and further reduce reliance on material-specific modeling.

Beyond spin, \gls*{RIXS} couples directly to charge, orbital, and lattice degrees of freedom \cite{Ament2011resonant,Mitrano2024exploring,deGroot2024resonant}, thus constituting a natural platform for entanglement metrology across multiple sectors \cite{shen2025witnessing,ren2024witnessing,Kwon2026intertwined} and for the identification of novel quantum phase transitions. Proposed \gls*{RIXS}-based protocols can already witness orbital entanglement \cite{shen2025witnessing,ren2024witnessing} and, when combined with \gls*{ARPES}, resolve entanglement of indistinguishable particles \cite{liu2025entanglement}. By extending entanglement metrology to time-resolved RIXS studies of optically driven quantum states \cite{Dean2016ultrafast,Mitrano2019Ultrafast,Parchenko2020orbital,Mazzone2021laser,Padma2025symmetry,Jost2026collective,merzoni2025photogenerated,Padma2026light}, our work also enables the use of entanglement as a figure of merit for photoinduced and coherent phases of matter \cite{Hales2023Witnessing,Baykusheva2023witnessing}.

\section*{Acknowledgments}
We thank P. Abbamonte, S. R. Clark, R. Fazio, M. Hafezi, T. Roscilde, S. Sachdev, A. O. Scheie, Q. Si, and A. Tennant for insightful discussions. We acknowledge technical support by F. Glerean. Experiments were primarily supported by the U.S.\ Department of Energy, Office of Basic Energy Sciences, Early Career Award Program, under Award No.\ DE-SC0022883. Theoretical and numerical work (Z.S., P.S., and Y.W.) was primarily supported by the U.S. Department of Energy, Office of Science, Basic Energy Sciences, under Early Career Award No.~DE-SC0024524. Work performed at Brookhaven National Laboratory (RIXS and RIXS data interpretation) was supported by the U.S.\ Department of Energy (DOE), Division of Materials Science, under Contract No.~DE-SC0012704. The dynamical DMRG simulations (D.B. and S.J.) were supported by the U.S. Department of Energy, Office of Science, Office of Basic Energy Sciences, under Award Number DE-SC0022311. S. F. R. T. acknowledges support by the U.S. Department of Energy, Office of Science, Office of Workforce Development for Teachers and Scientists, Office of Science Graduate Student Research (SCGSR) program. The SCGSR program is administered by the Oak Ridge Institute for Science and Education (ORISE) for the DOE. ORISE is managed by ORAU under contract number DE-SC0014664. All opinions expressed in this paper are the authors' and do not necessarily reflect the policies and views of DOE, ORAU, or ORISE. M. M. also acknowledges support by a Mercator Fellowship of the German Research Foundation (DFG) [Research Unit 5750, ``Optical Control of Quantum Materials" (OPTIMAL)]. V.M. was funded by a Harvard Quantum Initiative fellowship. P.S. is also partially supported by the Scialog Grant No. SA-QMI-2025-083a from Research Corporation for Science Advancement and Kevin Wells. The simulations used resources of the National Energy Research Scientific Computing Center, a U.S. Department of Energy Office of Science User Facility located at Lawrence Berkeley National Laboratory, operated under Contract No.~DE-AC02-05CH11231.

\section*{Data Availability}
The data supporting the findings of this study are provided in the paper and its Supplemental Material. Additional data are available from the corresponding authors upon reasonable request.

\section*{Appendix} 
\setcounter{section}{0}

\setcounter{figure}{0}
\renewcommand{\thefigure}{A\arabic{figure}}

\subsection{Sample growth and characterization}\label{ap:ex_growth}

High-quality Sr$_2$CuO$_3$ single crystals were grown by the traveling solvent floating-zone method. SrCO$_3$ and CuO (99.99\%) powders were mixed in their metal ratio, ground, and calcined (24~h at 980~$^{\circ}$C and 48~h at 1050~$^{\circ}$C for feed-rod material; 950~$^{\circ}$C for Sr$_2$CuO$_3$ solvent material). The powders were reground, packed into a rubber tube, and hydrostatically pressed at 4200~kg/cm$^2$ to form feed rods (8~mm diameter, 25~cm length), which were sintered in air for 72~h at 1100~$^{\circ}$C. Crystal growth was performed in a floating-zone furnace with two ellipsoidal mirrors, and the feed and seed rods were counter-rotated to ensure homogeneous heating and mixing. Single-crystal rods were grown at 1~mm/h under 1~bar O$_2$, yielding rods of $\sim$8~mm diameter and $\sim$10~cm length. The rods were cut into mm-sized crystals and characterized by X-ray diffraction to verify the degree of crystallinity and rule out undesired secondary phases. The lattice parameters are $a = 3.496$ \AA, $b = 3.906$ \AA, and $c = 12.684$ \AA, with chains oriented along the $b$ direction.

\subsection{RIXS measurements}\label{ap:ex_methods}

We conducted high-resolution \gls*{RIXS} measurements at beamline 2-ID of the National Synchrotron Light Source II, Brookhaven National Laboratory. Single-crystal samples were oriented using Laue diffraction, cleaved in rough vacuum, and immediately transferred to ultra-high vacuum for the \gls*{RIXS} measurements. The incident X-rays were tuned to resonance with the Cu $L_3$-edge peak (932.5~eV). The X-rays were $\pi$-polarized to maximize sensitivity to the spin fluctuations. The $ab$ plane was oriented in the scattering plane and the $c$-axis perpendicular to it. The scattering angle was fixed at $2\Theta = 150^\circ$ to maximize the total momentum transfer, while the sample angle $\theta$ was varied to control the momentum transfer along the $b$-axis chain direction. Given the \gls*{1D} character of the chains, we neglect dispersion along the $H$ and $L$ directions. We verified that the cleaved surfaces remained unoxidized by monitoring the \gls*{XAS} spectrum and, prior to each measurement, set the incident energy to the \gls*{XAS} absorption peak to ensure consistency among \gls*{RIXS} spectra and rule out energy drifts. The combined \gls*{RIXS} energy resolution was 35~meV. Further details of the \gls*{RIXS} data processing are discussed in the Supplemental Material Sec. S2 \cite{Supplementary}.

\subsection{Numerical Methods}\label{app:numerical_methods}

\subsubsection{Exact Diagonalization Calculations of RIXS Intensities}

We performed exact diagonalization calculations of the RIXS intensity, $\mathcal{I}_{\text{RIXS}}$, on chains of up to $L=16$ sites with periodic boundary conditions. We consider the \gls*{1D} \gls*{EHM} Hamiltonian at half filling,
\begin{equation}
\hat{H} = -t_h \sum_{\langle ij\rangle,\sigma} \left( \hat{c}^\dagger_{i\sigma}\hat{c}_{j\sigma} + \text{H.c.} \right) + U \sum_i \hat{n}_{i\uparrow}\hat{n}_{i\downarrow} + V \sum_{\langle ij\rangle} \hat{n}_i \hat{n}_j ,
\end{equation}
with nearest-neighbor hopping $t_h\simeq 560~\mathrm{meV}$, on-site interaction $U=8t_h$, and nearest-neighbor interaction $V=-t_h$. We obtained the full many-body spectrum and eigenstates using Lanczos methods to enable the calculation of dynamical correlation functions. 

The \gls*{RIXS} intensity is defined as~\cite{Ament2011resonant}
\begin{equation}
\begin{aligned}
    & \mathcal{I_{\text{RIXS}}}(q,\hbar\omega_{\rm in},\hbar\omega, \hat{e}_{\rm in}, \hat{e}_{\rm out}) \\& \qquad =  \frac{1}{\pi}\text{Im}\bra{\Psi_{\rm f}}\frac{1}{\hat{H}-E_G-\hbar\omega-i0^+}\ket{\Psi_{\rm f}}\,,
\end{aligned}
\end{equation}
where the final-state wavefunction is given by
\begin{equation}
\begin{aligned}
    \ket{\Psi_{\rm f}}  = \sum_{j} e^{i\textit{\textbf{q}} \cdot \textit{\textbf{r}}_j} \cdot \Bigl[  \mathcal{D}_{j\hat{e}_{\rm out}}^\dagger  \frac{1}{\hat{H}'-E_G-\hbar\omega_{\rm in}-i\Gamma}\mathcal{D}_{j\hat{e}_{\rm in}}\ket{G} \Bigr] \,.
\end{aligned}
\end{equation}
In these equations, $\textit{\textbf{q}}$ denotes the momentum transfer, $\hbar\omega_{\rm in}$ the incident energy, $\hbar\omega$ the energy loss, and $\hat{e}_{\rm in}$ ($\hat{e}_{\rm out}$) the incident (emitted) X-ray polarization. $E_G$ denotes the ground-state energy and $j$ the excitation site. The intermediate-state Hamiltonian $\hat{H}'$ includes the spin-orbit-coupled core hole and its interaction with valence electrons. The intermediate-state lifetime is defined by the phenomenological parameter $1/\Gamma$, which is incorporated by introducing a Lorentzian broadening $\Gamma$ in frequency space. We varied $\Gamma$ between $0.5t_h$ and $10t_h$ to assess its effect on spectral line shapes. Experimentally, the inverse lifetime $\Gamma$ is found to be $\sim 0.5t_h$ from the \gls*{XAS} spectrum (Supplemental Material Sec.~S1 \cite{Supplementary}), comparable to other cuprates \cite{Dean2012spin}.

The dipole operators $\mathcal{D}_{j\hat{e}_{\rm in}}$ ($\mathcal{D}_{j\hat{e}_{\rm out}}^\dagger$) represent orbital- and polarization-specific transitions between the core and valence bands by absorbing (emitting) an X-ray photon at a specific absorption edge at site $j$. For Cu $L_3$-edge \gls*{RIXS} in a $d^9$ system, these operators correspond to transitions between the $p_{3/2}$ core level and the $3d_{x^2-y^2}$ valence state and can be expressed as $\mathcal{D}_{j\hat{e}} = \sum_{\alpha,\sigma} A_{\alpha\sigma}^{\hat{e}} p_{j\alpha\sigma}^\dagger d_{j\sigma}$, where $p_{j\alpha\sigma}^{\dagger}$ creates a hole with spin $\sigma$ in orbital $\alpha$ (in the $LS$ representation), and $d_{j\sigma}$ annihilates a hole with spin $\sigma$, at site $j$. The dipole matrix element $A_{\alpha\sigma}^{\hat{e}} = \langle p_{\alpha\sigma}|\hat{e} \cdot \textit{\textbf{r}} | d_\sigma \rangle$ contains the geometric overlap between the absorbed (scattered) X-ray polarization, $\hat{e}_{\rm in}$ ($\hat{e}_{\rm out}$), and the atomic core and valence orbitals, which depends on the scattering geometry. To facilitate calculation, \gls*{ED} is used to calculate the fundamental excitation intensities, omitting the individual dipole matrix elements, $A_{\alpha\sigma}^{\hat{e}}$. The total scattering matrix element, $\mathcal{M}$, combining all contributions from individual $A_{\alpha\sigma}^{\hat{e}}$, is separately computed from an atomic model~\cite{Wang2019}. The fundamental excitation intensities and total scattering matrix elements are then multiplied together to give the \gls*{RIXS} intensity for each polarization- and spin-specific excitation channel \cite{Ament2009theoretical, jia2016using} (Supplemental Material Sec.~S3~\cite{Supplementary}). 

The total \gls*{RIXS} intensity, $\mathcal{I}_{\text{RIXS}}$, is the sum of two processes: a spin-flip process which reverses a spin between the initial and final states due to the core-hole spin-orbit coupling in the intermediate state $H'$, and a spin-conserving process which preserves the total spin of the system. The spin-conserving ($\mathcal{I}_{\mathrm{SC}}$) and spin-flip ($\mathcal{I}_{\mathrm{SF}}$) \gls*{RIXS} intensities are each computed separately by theoretically selecting the appropriate dipole transition operators and final matrix elements. The polarization configuration, together with the sample and scattering angles, sets the selection rules and thus determines whether spin-conserving or spin-flip excitations dominate the spectra. For the scattering geometry used here, both spin-conserving and spin-flip excitations occur in the $\pi$-$\pi$ scattering channel. In the scattering geometry used for typical \gls*{2D} cuprate measurements, the spin-flip excitations occur in channels which rotate the linear X-ray polarization by 90$^\circ$, while the spin-conserving excitations occur in the polarization-conserving channels (Supplemental Material Sec. S3A \cite{Supplementary}).

\subsubsection{Ground-State DMRG}

To determine $S(q,\hbar\omega)$ and the \gls*{QFI} of the Hubbard chain, we use DMRG to compute the ground state $|G\rangle$ of a 128-site \gls*{EHM} with open boundary conditions. Upon convergence, the resulting ground state is numerically exact within the DMRG representation~\cite{schollwock2005density,white1992density}. We then simulate the wavefunction following a local spin excitation and evaluate the equal-time correlation function in real space. The zero-temperature \gls*{QFI} density is then obtained by computing an appropriately normalized Fourier transform of this correlation function. 

\subsubsection{Numerical results}

\begin{figure}
    \centering
    \includegraphics[width=\linewidth]{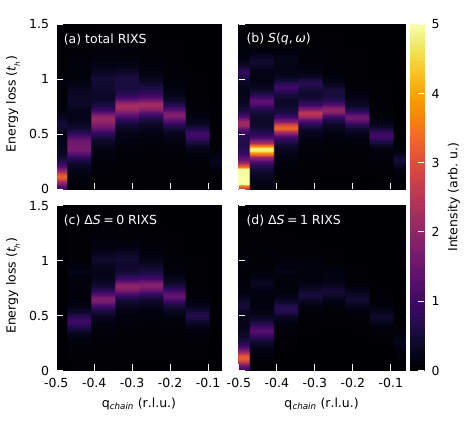}
    \caption{Numerical calculations of a \gls*{1D} \gls*{EHM}. (a) Total calculated \gls*{RIXS} intensity. (b) Spin structure factor. (c) Spin-conserving 2-spinon \gls*{RIXS} excitation intensity. (d) Spin-flip 2-spinon \gls*{RIXS} excitation intensity.}
    \label{fig:theory_RIXS_Sqw}
\end{figure}

The \gls*{ED}-simulated \gls*{RIXS} (at resonance energy) and \gls*{DMRG}-simulated $S(q,\hbar\omega)$ spectra are shown in Figs.~\ref{fig:theory_RIXS_Sqw}(a) and (b), respectively. Although the peak positions are similar, the intensities clearly have different momentum dependencies, primarily due to the mixed contributions to \gls*{RIXS} from spin-conserving and spin-flip 2-spinon excitations, which are shown individually in Figs.~\ref{fig:theory_RIXS_Sqw}(c) and (d), respectively. The spin-conserving excitations are strongest near $|q_{\text{chain}}| = 0.25$ r.l.u. In contrast, at $|q_{\text{chain}}|=0.5$ r.l.u., only $\Delta S = 1$ excitations are possible, and the \gls*{RIXS} spectrum is directly proportional to $S(q,\hbar\omega)$. The total projection factor $\mathcal{N}(q, \Gamma)$ in Eq.~\eqref{eq:norm} is determined by dividing the calculated $S(q,\hbar\omega)$ by the calculated total \gls*{RIXS} intensity.

\begin{figure}[H]
    \centering
    \includegraphics[width=0.9\linewidth]{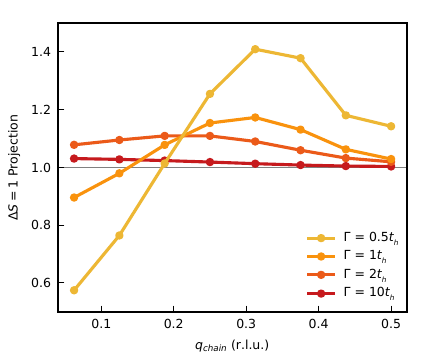}
    \caption{Projection of spin-flip ($\Delta S=1$) RIXS onto $S(q,\hbar\omega)$ as a function of inverse core-hole lifetime $\Gamma$.}
    \label{fig:corehole_lifetime_correction}
\end{figure}

Since for many materials it is possible to experimentally distinguish spin-conserving and spin-flip \gls*{RIXS} channels through their polarization dependence, we also calculate the projection factor from $\Delta S=1$~\gls*{RIXS} to the structure factor $S(q,\hbar\omega)$ for different values of the core-hole lifetime, $1/\Gamma$, by taking the ratio between the $\Delta S=1$ \gls*{RIXS} and the structure factor $S(q,\hbar\omega)$ (Fig. \ref{fig:corehole_lifetime_correction}). As expected, in the limit of an infinitely short intermediate state lifetime ($\Gamma \rightarrow \infty$), the projection factor approaches 1 for all momenta and the spin-flip \gls*{RIXS} channel reflects the structure factor $S(q,\hbar\omega)$. As the inverse lifetime $\Gamma$ decreases, the spectral weight is redistributed across different $\textbf{q}$. In experiments at the Cu $L_3$-edge, the core-hole inverse lifetime $\Gamma \sim 0.5 t_h$, estimated from the peak width of the \gls{XAS} (Supplemental Material Sec.~S1 \cite{Supplementary}). We use this value of $\Gamma = 0.5 t_h$ to obtain the total projection factor shown in Fig. \ref{fig:RIXS_corrections}(c) of the main text.

\bibliography{SCO_QFI_RIXS.bib}

\end{document}

\setlength{\parindent}{2ex}

\vspace{-4cm}

\renewcommand{\thesection}{S\arabic{section}}

\title[Article Title]{\hspace{1.cm}Supplementary Information for:\\ Observation of multipartite spin entanglement in a cuprate chain}

\author{S. F. R. TenHuisen}
\affiliation{Department of Physics, Harvard University, Cambridge, Massachusetts 02138, USA}
\affiliation{John A. Paulson School of Engineering and Applied Sciences, Harvard University, Cambridge, Massachusetts 02138, USA}

\author{Z. Shen}
\affiliation{Department of Chemistry, Emory University, Atlanta, GA 30322, USA}

\author{V. Bhartiya}
\affiliation{NSLS-II, Brookhaven National Laboratory, Upton, New York 11973, USA}

\author{V. Menon}
\affiliation{Department of Physics, Harvard University, Cambridge, Massachusetts 02138, USA}

\author{P. Sharma}
\affiliation{Department of Chemistry, Emory University, Atlanta, GA 30322, USA}

\author{H. Padma}
\affiliation{Department of Physics, Harvard University, Cambridge, Massachusetts 02138, USA}

\author{Z. Guan}
\affiliation{Department of Physics, Harvard University, Cambridge, Massachusetts 02138, USA}

\author{W. He}
\author{M. K. Lajer}
\affiliation{Condensed Matter Physics and Materials Science Department, Brookhaven National Laboratory, Upton, New York 11973, USA}

\author{J. Li}
\affiliation{NSLS-II, Brookhaven National Laboratory, Upton, New York 11973, USA}

\author{D. Banerjee}
\affiliation{Department of Physics and Astronomy, The University of Tennessee, Knoxville, Tennessee 37996, USA}
\affiliation{Institute for Advanced Materials and Manufacturing, University of Tennessee, Knoxville, Tennessee 37996, USA}

\author{J. Pelliciari}
\affiliation{NSLS-II, Brookhaven National Laboratory, Upton, New York 11973, USA}

\author{I. A. Zaliznyak}
\affiliation{Condensed Matter Physics and Materials Science Department, Brookhaven National Laboratory, Upton, New York 11973, USA}

\author{G. D. Gu}
\affiliation{Condensed Matter Physics and Materials Science Department, Brookhaven National Laboratory, Upton, New York 11973, USA}

\author{M. D. Lukin}
\affiliation{Department of Physics, Harvard University, Cambridge, Massachusetts 02138, USA}

\author{S. Johnston}
\affiliation{Department of Physics and Astronomy, The University of Tennessee, Knoxville, Tennessee 37996, USA}
\affiliation{Institute for Advanced Materials and Manufacturing, University of Tennessee, Knoxville, Tennessee 37996, USA}

\author{M. P. M. Dean}
\affiliation{Condensed Matter Physics and Materials Science Department, Brookhaven National Laboratory, Upton, New York 11973, USA}

\author{V. Bisogni}
\affiliation{NSLS-II, Brookhaven National Laboratory, Upton, New York 11973, USA}

\author{Y. Wang}\email[]{yao.wang@emory.edu}
\affiliation{Department of Chemistry, Emory University, Atlanta, GA 30322, USA}

\author{M. Mitrano}\email[]{mmitrano@g.harvard.edu}
\affiliation{Department of Physics, Harvard University, Cambridge, Massachusetts 02138, USA}

\date{\today}

\maketitle

\vspace{-10mm}
\tableofcontents

\newpage

\section{XAS}\label{sec:XAS}

\begin{figure}[h]
\centering
\includegraphics[width=0.55\linewidth]{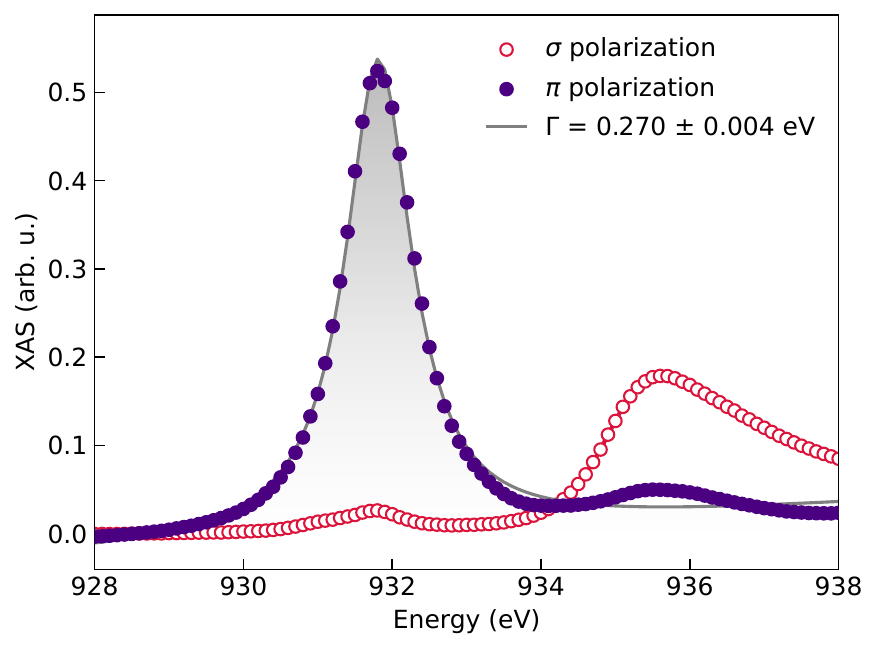}
\caption{\Gls{XAS} spectra measured with $\pi$-polarized (purple solid markers) and $\sigma$-polarized (red open markers) incident X-rays at an incident angle of $\theta = 63^\circ$. A Lorentzian fit to the $\pi$-polarized spectrum (gray solid line) yields a half-width at half-maximum of $\Gamma = 0.27$~eV.}
\label{fig:XAS}
\end{figure}

\Gls{XAS} was measured using an \gls{APD} placed in the scattering plane at $2\Theta = 150^\circ$. Representative \gls{XAS} spectra are shown in Fig.~\ref{fig:XAS} for $\pi$-polarized (purple solid markers) and $\sigma$-polarized (red open markers) incident light. Sr$_2$CuO$_3$ exhibits a single, strongly dichroic peak at the Cu L$_3$ edge ($\sim 932$~eV), with strong absorption for light polarized within the CuO$_4$ plaquettes ($\pi$ polarization) and negligible absorption for light polarized perpendicular to the plaquettes ($\sigma$ polarization). Strong XAS dichroism indicates high-quality samples, whereas over- and under-oxidized samples show appreciable additional peaks in both polarization channels.

We estimate the inverse core-hole lifetime $\Gamma$ from the half-width at half-maximum of the main absorption edge. A Lorentzian fit yields $\Gamma = 0.27$~eV for Sr$_2$CuO$_3$, consistent with typical values for copper oxides~\cite{Dean2012spin}.

\section{RIXS data processing}\label{sec:processing}
Here we describe the data processing for the RIXS datasets presented in the main text. In order to determine the \gls*{QFI}, the magnetic intensities must be accurately determined in absolute units. Thus, we correct the data for self-absorption effects, which suppress scattering intensity for grazing-exit conditions (Sec.~\ref{sec:self_absorption}). We then benchmark the intensities by comparing the $d_{3z^2-r^2}$ orbital intensity to that expected by an atomic model calculation (Sec.~\ref{sec:dd_norm}), correcting for variations in beam footprint and collection efficiency at each angle. Finally, we fit and subtract all non-magnetic spectral weight from the \gls*{RIXS} spectra (Sec.~\ref{sec:fitting}). 

\subsection{Self-absorption correction}\label{sec:self_absorption}
The \gls*{RIXS} data are corrected for self-absorption effects, accounting for the energy dependence of the emitted light and the sample geometry, and assuming an infinitely thick sample relative to the penetration depth of the incident light according to the equations in Ref. \cite{Wang2020}. Since the penetration depth at the Cu L$_3$-edge is of order 50 nm and far smaller than the sample thickness, this approximation is accurate. 
The correction factor, $C$, is given by
\begin{equation}
    C = \frac{1}{1+u\cdot t},
\end{equation}

where $u$ accounts for the angle-dependent path lengths, according to 
\begin{equation}
    u = \frac{\sin{\theta}}{\sin{(2\Theta-\theta)}},
\end{equation}
and $t$ accounts for the difference in absorption along different crystallographic directions, according to
\begin{equation}
    t = \frac{\alpha_0 + \alpha_{\epsilon}(E_{out},\theta,2\Theta)}{\alpha_0+ \alpha_{\epsilon}(E_{in},\theta,2\Theta)}.
\end{equation}
$\alpha_0$ is the X-ray absorption below the absorption edge; $\alpha_{\epsilon}$ is the X-ray absorption at a given X-ray energy and along a given crystallographic direction, given by 
\begin{equation}
     \alpha_{\epsilon}(E,\theta, 2\Theta) = \alpha_{\parallel b}(E) \sin^2{(2\Theta-\theta)} + \alpha_{\parallel c}(E) \cos^2{(2\Theta - \theta)}.
\end{equation}

We assume the outgoing light is $\pi$-polarized, as single-site cluster calculations indicate that in the absence of orbital excitations, the cross-section for $\sigma$-polarized light is zero for the scattering geometry used here (Sec. \ref{sec:matrix_elements}). Self-absorption effects are primarily important for outgoing energies near the main \gls*{XAS} absorption edge, and thus mainly affect the low-energy-loss region of the spectrum, below the orbital excitation energy.

\subsection{Orbital normalization}\label{sec:dd_norm}

\begin{figure}
    \centering
    \includegraphics[width=0.6\linewidth]{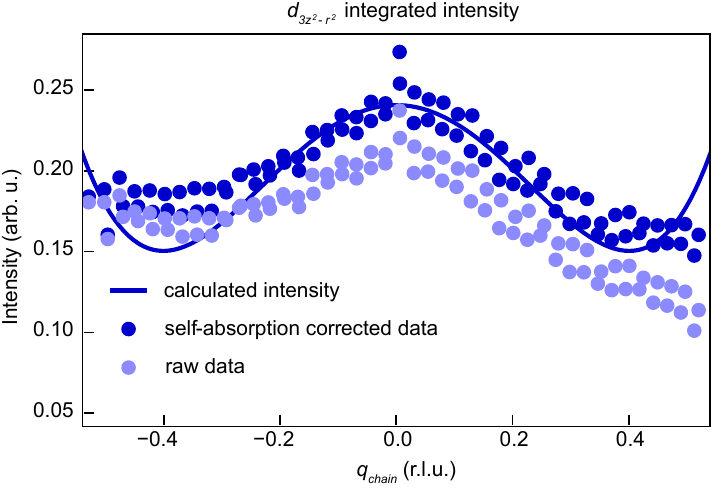}
    \caption{Integrated intensity of the $d_{3z^2-r^2}$ orbital. The calculated intensity is shown as a solid dark blue line. The integrated measured intensity before self-absorption is identified by light blue circles, and the integrated measured intensity after correcting for self-absorption effects by dark blue circles. Raw and self-absorption-corrected data in this figure contain independent datasets from the two RIXS detectors of the SIX beamline.}
    \label{fig:dd_normalization}
\end{figure}

To correct for angle-dependent variations in beam footprint and collection efficiency, we normalize the \gls*{RIXS} spectra at each angle to the calculated intensity of the $d_{3z^2-r^2}$ orbital, after correcting for energy- and angle-dependent self-absorption. We use only the $d_{3z^2-r^2}$ orbital because it points perpendicular to the CuO$_4$ chains and is therefore well localized. The remaining orbitals have appreciable overlap along the chain direction and, due to the \gls{1D} interactions in Sr$_2$CuO$_3$, undergo spin-orbital fractionalization. Their excitations are consequently dispersive along the chains and are not well described by local models~\cite{Wohlfeld2013microscopic, Schlappa2012spin}. 

We calculate the $d_{3z^2-r^2}$ orbital intensity using an atomic model of a single Cu d$^9$ site in EDRIXS \cite{Wang2019}, in the same way the matrix elements in Sec. \ref{sec:matrix_elements} were computed. We consider a square-planar crystal field splitting, set to match the energies of the measured orbital excitations, with the $d_{3z^2-r^2}$ orbital lying at 3 eV energy loss in the \gls*{RIXS} spectra, the $d_{xz}$ and $d_{yz}$ orbitals at 2.2 eV energy loss, and the $d_{xy}$ orbital at 1.8 eV energy loss. The \gls*{RIXS} intensity is summed over both spin-conserving and spin-flip excitations to the $d_{3z^2-r^2}$ and over both $\sigma$ and $\pi$ outgoing polarizations. 

The experimentally measured $d_{3z^2-r^2}$ intensity (light blue circles), determined by integrating the \gls{RIXS} spectra from 2.85 to 3.5 eV to avoid contributions from nearby dispersive orbiton excitations, agrees well with the expected trend from the atomic model calculations (solid blue line) [Fig. \ref{fig:dd_normalization}]. After applying the self-absorption correction (dark blue circles), the agreement is further improved, with some variations only at large momentum transfers, where either the incoming X-rays or outgoing X-rays are quite grazing, and thus the alignment of the sample more strongly affects the collection efficiency.

\subsection{Fitting and subtracting non-magnetic features}\label{sec:fitting}

\begin{figure}
    \centering
    \includegraphics[width=\linewidth]{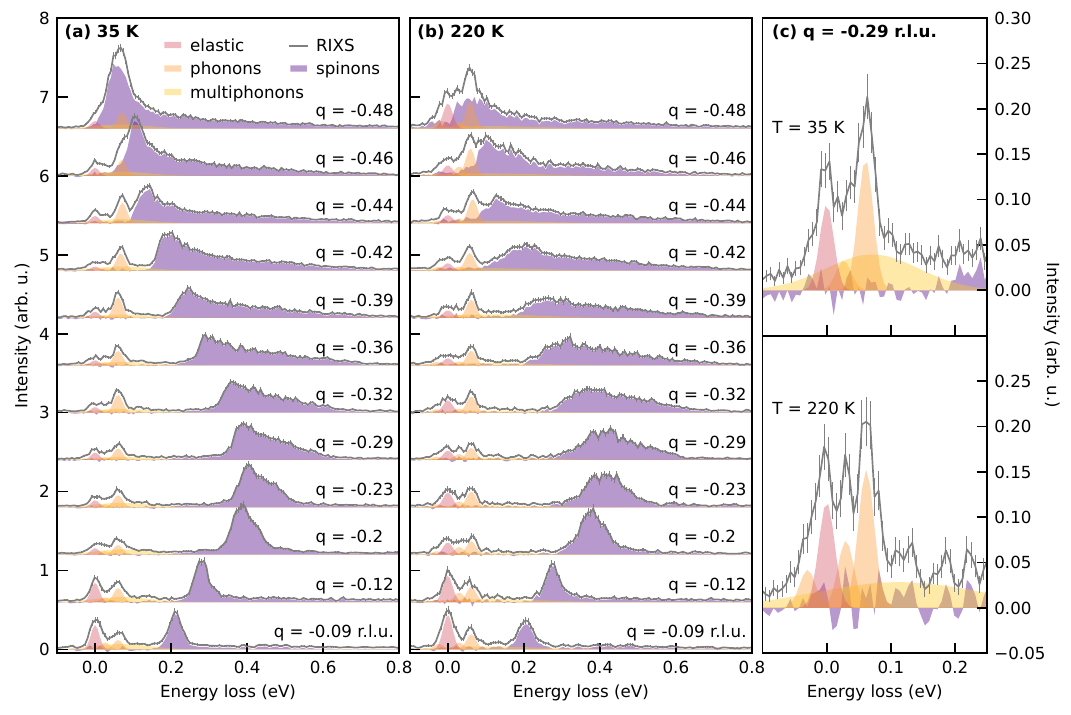}
    \caption{RIXS data with low-energy elastic and phonon fits at (a) 35 K and (b) 220 K. (c) Comparison of low-energy spectra and fit components for $q = -0.29$ r.l.u. at 35 K and 220 K. Experimentally measured RIXS spectra are shown in gray. The fitted quasielastic contribution is shown in pink, phonons in orange, and a broad multiphonon component is shown in yellow. The remaining spectral weight, attributed to spinon excitations, is shown in purple. }
    \label{fig:fits_all}
\end{figure}

To isolate the magnetic contributions to \gls*{RIXS}, we fit and subtract the quasielastic line and phonons. We model the \gls{RIXS} spectra as containing four low-energy features: a quasielastic peak at zero energy loss, two phonons with energies below 100 meV, and a multiphonon contribution at higher energy. We assume all peaks are Gaussian. Prior to fitting, we subtract the average value of the \gls{RIXS} spectrum between -1 and -0.15 eV on the energy gain side from each \gls{RIXS} spectrum, which constitutes a weak and constant background contribution. We do not include any additional background in the fitting model. 

At most momenta ($0.12 \lesssim |q| \lesssim 0.4$ r.l.u.), these non-magnetic features are energetically well-separated from the 2-spinon continuum. The phonons are visibly nondispersive, as expected for optical phonons. We simultaneously fit these \gls*{RIXS} spectra over the energy loss region from -0.3 to 0.18~eV: the energy of the phonon peaks and the peak widths are shared global free parameters, optimized over all spectra, while the amplitudes of each peak are independent in each spectrum. From these fits, we determine that the phonons are at 26 meV and 61 meV, respectively. At 35 K, the quasielastic peak is determined to have a Gaussian width $\sigma = 13$~meV, and the phonons a Gaussian peak width of $\sigma = 14$~meV. We use these fit parameters to constrain the fit parameters for the remaining spectra where the 2-spinon continuum overlaps with the other low-energy excitations. 

For the data at 220 K, we additionally include a small peak on the energy gain side, capturing weak anti-Stokes phonon spectral weight. The peak widths at 220 K are found to be slightly broader, with a width of $\sigma = 14.7$ meV, likely due to the thermal activation of additional low energy phonon modes at higher temperature. The spectra and their fits are shown in Fig.~\ref{fig:fits_all}. 

\section{Spin-flip and spin-conserving processes}\label{sec:S1S0}
The total \gls*{RIXS} intensities shown in Figs. 2(b) and A1 of the main text depend both on fundamental material properties and on the \gls*{RIXS} total scattering matrix elements, which depend on the experimental geometry. The total \gls*{RIXS} intensity is a sum of the intensities of the spin-flip ($\Delta S=1$) and spin-conserving ($\Delta S=0$) channels. The intensity of each channel can be expressed as a product of a geometrical total scattering matrix element and the fundamental scattering intensity, which can then be calculated separately. Thus, the total magnetic \gls*{RIXS} intensity for each polarization channel can be expressed as:  
\begin{equation}
    \begin{aligned}
        \mathcal{I}_{\text{RIXS}}&(\textbf{q}, \hbar\omega_\text{in}, \hbar\omega, \hat{e}_\text{in},\hat{e}_\text{out}) \\ & = \mathcal{I}_{\text{SF}}(\textbf{q}, \hbar\omega_\text{in}, \hbar\omega, \hat{e}_\text{in},\hat{e}_\text{out}) + \mathcal{I}_{\text{SC}}(\textbf{q}, \hbar\omega_\text{in}, \hbar\omega, \hat{e}_\text{in},\hat{e}_\text{out}) \\
    & = \mathcal{M}_{\text{SF}}(\hat{e}_\text{in},\hat{e}_\text{out})I_{\text{SF}}(\textbf{q}, \hbar\omega_\text{in}, \hbar\omega) + \mathcal{M}_{\text{SC}}(\hat{e}_\text{in},\hat{e}_\text{out})I_{\text{SC}}(\textbf{q}, \hbar\omega_\text{in}, \hbar\omega) .
    \end{aligned}
\end{equation}
$\mathcal{I}_{\text{SF}}(\textbf{q}, \hbar\omega_\text{in}, \hbar\omega, \hat{e}_\text{in},\hat{e}_\text{out})$ ($\mathcal{I}_{\text{SC}}(\textbf{q}, \hbar\omega_\text{in}, \hbar\omega, \hat{e}_\text{in},\hat{e}_\text{out})$) is the total spin-flip (spin-conserving) \gls*{RIXS} intensity. $\mathcal{M}_{\text{SF}}(\hat{e}_\text{in},\hat{e}_\text{out})$ ($\mathcal{M}_{\text{SC}}(\hat{e}_\text{in},\hat{e}_\text{out})$) is the geometrical total scattering matrix element for spin-flip (spin-conserving) processes for incident polarization $\hat{e}_\text{in}$ and scattered polarization $\hat{e}_\text{out}$; $I_{\text{SF}}(\textbf{q}, \hbar\omega_\text{in}, \hbar\omega)$ ($I_{\text{SC}}(\textbf{q}, \hbar\omega_\text{in}, \hbar\omega)$) is the fundamental bare intensity of spin-flip (spin-conserving) excitation for momentum transfer $\textbf{q}$, incident X-ray energy $\hbar\omega_\text{in}$, and energy loss $\hbar\omega$. We now discuss the separate contributions of the total scattering matrix elements (Sec.~\ref{sec:matrix_elements}) and fundamental excitation intensities (Sec. \ref{sec:bare_intensities}) to the total magnetic \gls*{RIXS} scattering. 

\subsection{RIXS total scattering matrix elements}\label{sec:matrix_elements}

The total scattering matrix elements were computed for an atomic model using EDRIXS \cite{Wang2019} by calculating the X-ray transition amplitudes between specific states for a single $d^9$ Cu site. We assumed a tetragonal crystal field, with the $d_{3z^2-r^2}$ orbital having the lowest energy, followed by the degenerate $d_{xz}$ and $d_{yz}$ orbitals, then the $d_{xy}$ orbital, and the half-filled $d_{x^2-y^2}$ orbital as the valence orbital. We specified a small magnetic field to account for the exchange field in the real sample and break the degeneracy of spin-up and spin-down states. We diagonalized the resulting Hamiltonian to determine the eigenvectors and then calculated the transition amplitudes, $F_{fg} = \langle f | \mathcal{D} \mathcal{D}^\dagger |g \rangle $, from the ground state $|g\rangle$ to each orbital configuration $|f\rangle$. For spin-conserving excitations, the final orbital and spin configuration is the same as the initial configuration, while for spin-flip excitations, the final configuration has the same orbital occupancies as the initial state but with a reversed spin in the valence $d_{x^2-y^2}$ orbital. 

We then determined the total scattering matrix elements for spin-flip ($\mathcal{M}_{\text{SF}}(\hat{e}_\text{in},\hat{e}_\text{out})$) and spin-conserving ($\mathcal{M}_{\text{SC}}(\hat{e}_\text{in},\hat{e}_\text{out})$) excitations to and from the ground-state $d_{x^2-y^2}$ orbital by determining the polarization vectors for the incident and scattered X-rays. The total scattering matrix element, $\mathcal{M}_{fg}(\hat{e}_\text{in},\hat{e}_\text{out})$, is given by 
\begin{equation}
    \mathcal{M}_{fg}(\hat{e}_\text{in},\hat{e}_\text{out}) = | \hat{e}_\text{out} \cdot F_{fg} \cdot \hat{e}_\text{in} |^2
\end{equation}
where $\hat{e}_{\mathrm{out,in}}$ are the outgoing and incident polarization vectors. $\mathcal{M}$ captures the geometric and polarization dependences which can alternatively be encoded in individual dipole matrix elements, $A_{\alpha\sigma}^{\hat{e}}$, in the \gls*{RIXS} intensity calculations. We use $\pi$ to denote light polarized in the scattering plane, while $\sigma$ denotes light polarized perpendicular to the scattering plane.

\begin{figure}
    \centering
    \includegraphics[width=0.45\linewidth]{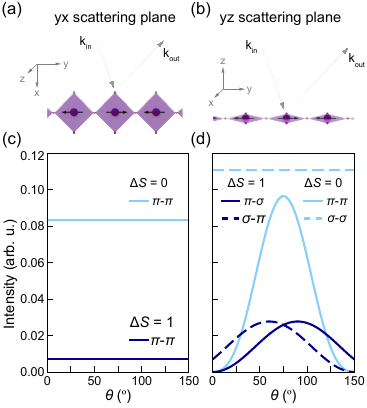}
    \caption{Total scattering matrix elements versus sample angle $\theta$ at fixed scattering angle $2\Theta=150^\circ$ for each polarization channel in (a), (c) the $xy$ scattering plane and (b),(d) the $yz$ scattering plane. Spin-flip ($\Delta S=1$) and spin-conserving ($\Delta S=0$) channels are shown in dark and light blue, respectively. Solid (dashed) lines denote incident $\pi$ ($\sigma$) polarization.}
    \label{fig:SF_NSF}
\end{figure}

The scattering geometry influences the total scattering matrix elements [Fig.~\ref{fig:SF_NSF}]. For the $xy$ scattering geometry [Fig.~\ref{fig:SF_NSF}(a)] imposed by the natural cleavage planes of Sr$_2$CuO$_3$, only matrix elements for $\pi$-polarized incident and emitted light are non-zero [Fig.~\ref{fig:SF_NSF}(c)]. As the $d_{x^2-y^2}$ orbital lies in the scattering plane, the overlap between the orbital and the light polarization is independent of the sample angle $\theta$, and thus both spin-conserving and spin-flip total scattering matrix elements are angle-independent. The total scattering matrix element for the spin-conserving channel is larger than the spin-flip channel for $2 \Theta = 150^\circ$. In contrast, for the $yz$ scattering geometry [Fig.~\ref{fig:SF_NSF}(b)] used to study most \gls*{2D} cuprates, all combinations of incident and emitted polarizations are possible, with the spin-conserving ($\Delta S=0$) channel corresponding to conserved polarization and the spin-flip ($\Delta S=1$) channel corresponds to flipping the polarization between absorption and emission [Fig.~\ref{fig:SF_NSF}(d)]. As the $d_{x^2-y^2}$ orbital is oriented perpendicular to the scattering plane, the overlap with $\pi$-polarized light is strongly dependent on the sample angle $\theta$. For $2 \Theta = 150^\circ$, the spin-flip channel ($\Delta S = 1$) dominates for \gls*{RIXS} measurements with $\pi$-polarized light near grazing-exit conditions. 

\subsection{Fundamental scattering intensities}\label{sec:bare_intensities}
\begin{figure}
    \centering
    \includegraphics[width=0.35\linewidth]{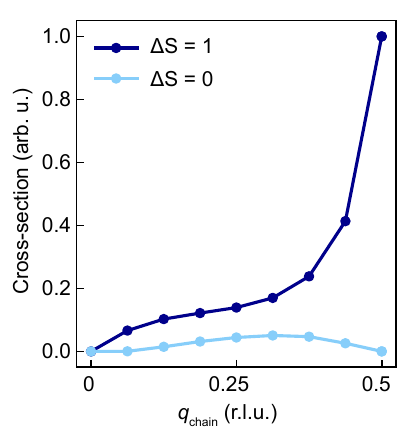}
    \caption{Integrated intensity of the fundamental scattering intensity for the spin-conserving ($\Delta S=0$, light blue) and spin-flip ($\Delta S=1$, dark blue) 2-spinon excitations in an \gls*{EHM}.}
    \label{fig:bare_intensities}
\end{figure}

We calculate the fundamental RIXS scattering intensities for the spin-conserving ($\Delta S=0$) and spin-flip ($\Delta S=1$) channels for an EHM chain with parameters describing Sr$_2$CuO$_3$, as discussed in Appendix C. Figure~\ref{fig:bare_intensities} shows the fundamental scattering intensities integrated over the energy-loss axis. The spin-flip intensity peaks at the zone boundary, $q_{\rm chain}=0.5$ r.l.u., and is stronger than the spin-conserving intensity at all momenta. The spin-conserving intensity instead peaks at intermediate momenta, around $q_{\rm chain}=0.3$ r.l.u., and vanishes at the zone boundary.

The material-dependent core-hole lifetime influences the relative contributions of the spin-flip and spin-conserving channels. As spin-flip excitations are created directly from the absorption and emission dipole processes without any intermediate state dynamics, $I_{SF}$ is independent of the core-hole lifetime, $\Gamma$. In contrast, spin-conserving excitations arise due to a rearrangement of spins during the intermediate state lifetime. The rearrangements of neighboring spins occur on a characteristic superexchange timescale of $\tau = \hbar / J$, where $J$ is the spin superexchange energy \cite{Bisogni2014femtosecond}. The longer the superexchange timescale is relative to the core-hole lifetime, the larger the contribution from $I_{\mathrm{SC}}$ becomes. In Sr$_2$CuO$_3$, $J = 250$ meV, thus the timescale for spin rearrangements can be estimated as $\tau_{S0} = \hbar/J = 2.5$ fs, while the core-hole lifetime is $\tau = \hbar/\Gamma \sim 2$ fs. Thus in Sr$_2$CuO$_3$ at the Cu $L_3$-edge we expect a non-negligible contribution of spin-conserving excitations to the overall \gls*{RIXS} intensity.

\section{Temperature dependence}\label{sec:Tdep}

\begin{figure}[H]
    \centering
    \includegraphics[width=0.9\linewidth]{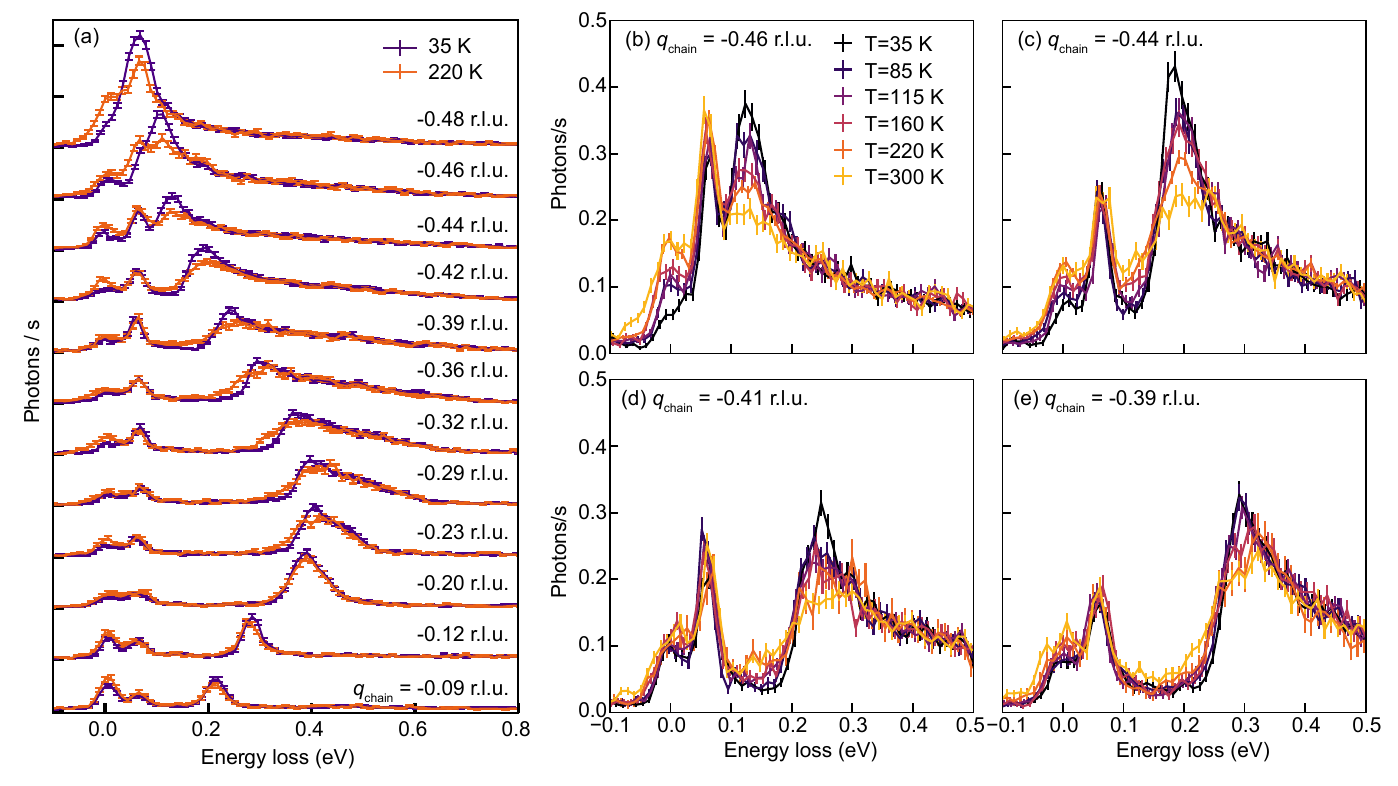}
    \caption{Temperature-dependent RIXS spectra. (a) Measured RIXS intensity as a function of momentum at 35 K (purple) and 220 K (orange). Measured RIXS intensity at fixed momentum for intermediate temperatures at (b) $q_{\mathrm{chain}} = -0.46$ r.l.u., (c) $-0.44$ r.l.u., (d) $-0.41$ r.l.u., and (e) $-0.39$ r.l.u.}
    \label{fig:Tdep}
\end{figure}

Fig.~\ref{fig:Tdep}(a) shows the full measured dispersion at 35 K (purple) and 220 K (orange). At each momentum, the low-energy quasielastic and phonon scattering increase slightly due to thermal occupation. The strong bond-stretching phonon at ~60 meV is minimally affected as this energy scale is above the thermal energy scale even at 220 K. Additionally, the 2-spinon continuum broadens prominently, particularly near the zone boundary, $|q_{\text{chain}}| = 0.5$. We also measured the RIXS intensity at intermediate temperatures at fixed momentum transfers of $q_{\text{chain}} = -0.46$, $-0.44$, $-0.41$, and $-0.39$ r.l.u., shown in Figs.~\ref{fig:Tdep}(b)-(e). For each momentum, we fixed the momentum transfer and measured the spectrum at temperatures from 35 to 300 K, allowing the sample to stabilize at each temperature. The spectra evolve continuously with temperature, with the largest changes in the two-spinon intensity near the zone boundary, $|q_{\text{chain}}| = 0.5$ r.l.u.

\newpage
\section{Impact of $U$ on spin dynamical structure factor}\label{sec:Udep}

\begin{figure}[H]
    \centering
    \includegraphics[width=0.55\linewidth]{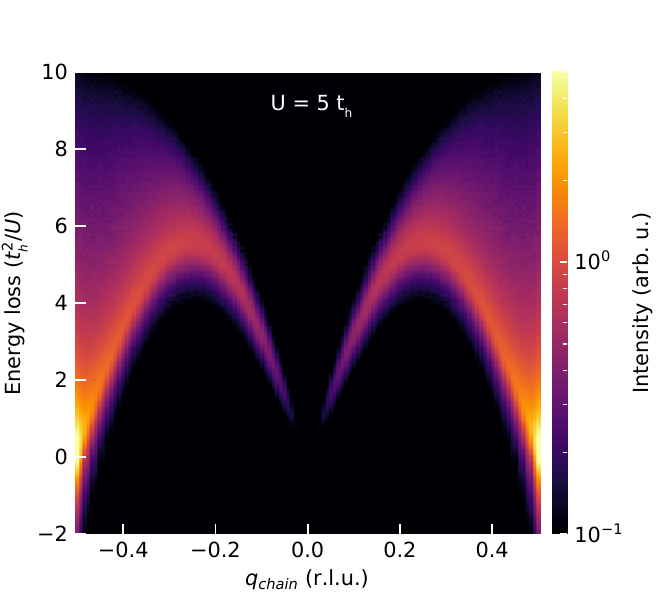}
    \caption{Calculation of the dynamical structure factor $S(q,\hbar\omega)$ for a Hubbard model with $U=5t_h$. }
    \label{fig:Sqw_U5}
\end{figure}

To test robustness against alternative parameterizations reported in the literature, we compare the spectrum obtained for the alternative parameterization $U \approx 5t_h$~\cite{Li2021particle} with expectations for the weak- and strong-coupling limits of the Hubbard model~\cite{Nocera2016magnetic} [Fig.~\ref{fig:Sqw_U5}]. The spectral weight is concentrated near the lower edge of the continuum, consistent with Heisenberg-like behavior and the ratio $t_h/U$. This supports interpreting the low-energy dynamics of Sr$_2$CuO$_3$ in terms of independent spin degrees of freedom.

\section{Additional notes on witnessing spin entanglement with QFI}\label{sec:supp_QFI_defns}

Here we expand upon the derivations underlying the entanglement analysis in Sec.~III of the main text. We provide additional details on the relations between the relevant equations, the normalization conditions required for entanglement metrology, and the sum rules obeyed by the dynamical spin structure factor. To improve readability, we repeat several key equations from the main text.

\subsection{Entanglement in pure and mixed states}
Because entanglement of a pure quantum state is a property of the wavefunction, we characterize it through the notion of $m$-producibility. A pure spin state $\ket{\Psi_{m\text{-}\rm{prod}}}$ is $m$-producible if it factorizes into non-overlapping partitions,
\begin{equation}\label{eq:spinState}
\ket{\Psi_{m\text{-}\rm{prod}}} = \ket{\Phi_1}\otimes \ket{\Phi_2}\otimes\cdots\otimes \ket{\Phi_\ell},
\end{equation}
where each $\ket{\Phi_a}$ is an irreducible many-body state defined on a partition $P_a$ with $|P_a| \le m$. The entanglement depth is then $m$ if the state is $m$-producible but not $(m-1)$-producible. The QFI can be computed simply as 
\begin{equation}
    f_Q = \frac{4}{N}\mathrm{Var}_\psi (\hat{O})= \frac{4}{N}(\langle \hat{O}^2\rangle - \langle \hat{O} \rangle^2).
\end{equation}

A mixed state is $m$-producible if the density matrix $\rho$ admits a convex decomposition \begin{equation}
\rho
=
\sum_\gamma w_\gamma
\ket{\Psi^{(\gamma)}_{m\text{-}\mathrm{prod}}}
\bra{\Psi^{(\gamma)}_{m\text{-}\mathrm{prod}}},
\end{equation}
where $w_\gamma \geq 0$, $\sum_\gamma w_\gamma=1$, and every
$\ket{\Psi^{(\gamma)}_{m\text{-}\mathrm{prod}}}$ is a pure
$m$-producible state, although its partition may differ between terms. In this case, the QFI is then defined as 
\begin{equation}
f_Q(\rho,\hat O)=\frac{2}{N}\sum_{n,n'}\frac{(p_n-p_{n'})^2}{p_n+p_{n'}}|\bra{n}\hat O\ket{n'}|^2,\label{eq:QFI_def_mixed}
\end{equation}
with $\rho=\sum_n p_n\ket{n}\bra{n}$ being the density matrix and $\hat O=\sum_{i=1}^N \hat O_i$ a sum of local Hermitian operators. For $\hat O_i$ with eigenvalues bounded between $\pm 1/2$
\begin{equation}
f_Q > m
\label{eq:entanglement_bound}
\end{equation}
certifies at least $(m+1)$-partite entanglement.

\subsection{Connection between entanglement and dynamical susceptibility}

Hauke \textit{et al.} showed that, for a thermal Gibbs state in equilibrium, $f_Q(\rho,\hat O)$ is experimentally accessible through a sum rule over the dynamical susceptibility~\cite{Hauke2016measuring}

\begin{equation}\label{eq:QFIchi}
    f_Q(\rho, \hat{O}) 
     = 4 \int_{0}^{\infty} \mathrm{d} (\hbar\omega) \tanh \left(\frac{\hbar \omega}{2k_BT}\right) \chi^{\prime \prime}_{\hat{O}\hat{O}}(\hbar \omega),
\end{equation}

where $\hbar \omega$ is the energy and  $\chi''_{\hat{O}\hat{O}}(\hbar \omega)$ is the imaginary part of the retarded correlation function \footnote{Note that here we use a site-normalized definition for $\chi''_{\hat{O}\hat{O}}(\hbar \omega)$,  whereas Ref.~\cite{Hauke2016measuring} uses an extensive definition that scales with the number of sites in the system.}
\begin{equation}\label{eq:chi}
	 \chi''_{\hat{O}\hat{O}}(\hbar \omega) = \frac{1}{\hbar N } \Im \bigg[i \int_0^{\infty} \!\frac{\mathrm{d}t}{\pi}~e^{i \omega t} \operatorname{Tr}\left(\rho\left[\hat{O}(t), \hat{O}(0)\right]\right)\bigg].
\end{equation}

Within this formalism, the $\tanh(\hbar \omega/2k_BT)$ factor suppresses classical low-energy thermal fluctuations from contributing to $f_Q(\rho, \hat{O})$. To apply this formalism to spin entanglement, we consider the spin operator $\hat{S}^\alpha$, where $\alpha \in\{x,y,z\}$ is the spin direction probed. We can then define a dynamical structure factor \footnote{The sign convention in the exponential is tied to the operator ordering: reversing the sign is equivalent to interchanging $\hat{S}^\alpha(t)$ and $\hat{S}^\alpha(0)$.}
 \begin{equation}\label{eq:general_dynamical_structure_factor}
S^{\alpha\beta}(q,\hbar\omega) = \frac{1}{2\pi N \hbar} \int_{-\infty}^{\infty}\mathrm{d}t\, e^{ i\omega t } 
\left\langle \hat{S}^\alpha_{-q}(t)\hat{S}^\beta_q(0)\right\rangle
\end{equation}
where angle brackets denote thermal averages $\langle \hat{A}\rangle_\rho = \operatorname{Tr}(\rho\hat{A})$, $q$ is the momentum transfer in reciprocal lattice units (r.l.u.) \footnote{This unit choice leads to the integral limits being $\pm 1/2$ rather than $\pm \pi$ later on.}, $r_i$ denotes the position of site $i$ in units of the lattice constant, and

\begin{equation}\label{eq:spinQop}
\hat S_q^\alpha
=
\sum_{j=1}^N e^{-i2\pi q r_j}\hat S_j^\alpha
\end{equation}

As such, the most general spin susceptibility is a $3\times3$ tensor of different spin directions $\chi_{\alpha\beta}$, where $\alpha,\beta\in\{x,y,z\}$. For an isotropic magnet, the diagonal components are equivalent, so it is sufficient to consider a single component
\begin{equation}\label{eq:dynamical_structure_factor}
S^\alpha(q,\hbar\omega) = \frac{1}{2\pi N \hbar} \int_{-\infty}^{\infty}\mathrm{d}t\, e^{ i\omega t } 
\left\langle \hat{S}^\alpha_{-q}(t)\hat{S}^\alpha_q(0)\right\rangle
\end{equation}

Through the fluctuation-dissipation theorem, the dynamical susceptibility can be connected to a generalized dynamical structure factor \footnote{ sometimes one sees an additional proportionality constant of $\pi$ in this relation. In our formalism,  $\pi$ instead appears in Eq.~\eqref{eq:dynamical_structure_factor}.}
\begin{equation}
\chi_\alpha^{\prime\prime}(q, \hbar \omega)
=
\left(
1-e^{- \hbar \omega/k_B T}
\right)
S^\alpha(q, \hbar \omega) .
\label{eq:chi_Sqom}
\end{equation}
Substituting Eq. \eqref{eq:chi_Sqom} into Eq. \eqref{eq:QFIchi} yields
\begin{equation}\label{eq:QFIIntegral}
f_Q(q,T) = 4 \int_{-\infty}^{\infty} \tanh^2\!\bigg(\frac{\hbar \omega}{2 k_B T}\bigg)\, S(q,\hbar\omega,T)\, d(\hbar\omega),
\end{equation}
as given in the main text. 

The single spin component of the dynamical structure factor is constrained by the sum rule for $S=1/2$ 
\begin{equation}
\int_{-1/2}^{1/2} \mathrm{d}q
\int_{-\infty}^{\infty} \mathrm{d} (\hbar \omega)\,
S^\alpha(q,\hbar\omega)
=
\frac{S(S+1)}{3}
=
\frac{1}{4} .
\label{eq:sum_rule}
\end{equation}
Note that this is one third of the typical total sum rule intensity, because only one of the three spin components is measured. This assumes spin-rotation invariance, so the total on-site moment $S(S+1)$ is shared equally among the three Cartesian coordinates. For the special case of $q \in \frac{1}{2}\mathbb{Z}$, $\hat S_q^\alpha = \hat S_{-q}^\alpha$ and we obtain the QFI density  corresponding to the spin operator 
\begin{equation}\label{eq:QFI_RIXS}
    \begin{aligned}
        f_Q(q, T)
        ={}& 4 \int_0^\infty \mathrm{d}(\hbar\omega)\,
        \tanh\left(\frac{\hbar \omega}{2k_BT}\right) \times \left(1-e^{-\hbar \omega/k_BT}\right)
        S^\alpha(q, \hbar\omega) \\
        ={}& 4 \int_{-\infty}^{\infty} \mathrm{d}(\hbar \omega)\,
        \tanh^2\left(\frac{\hbar\omega}{2k_BT}\right)
        S^\alpha(q,\hbar\omega).
    \end{aligned}
\end{equation}

\subsection{Generalization to arbitrary $\mathrm{q}$}
A QFI generator must be Hermitian. The Fourier component $\hat{S}_q^\alpha$ is itself Hermitian when $q=-q$ modulo a reciprocal lattice vector. In the one-dimensional
Brillouin zone used here, the important examples are
\begin{align}
    q&=0,
    &
    \hat{S}_0^\alpha&=\sum_j\hat{S}_j^\alpha,
    \\
    q&=\tfrac12,
    &
    \hat{S}_{1/2}^\alpha&=\sum_j(-1)^{r_j}\hat{S}_j^\alpha.
\end{align}
At the antiferromagnetic wavevector \(q=1/2\), each local term still has spectral width one, so the entanglement criterion Eq.~\eqref{eq:entanglement_bound} applies directly.

The construction can be generalized to witness entanglement at arbitrary
$q$ using the method described in Ref.~\cite{ren2024witnessing}. While $\hat S_q^\alpha$ is not generally Hermitian, it can be separated into two components that are themselves Hermitian at every $q$:

\begin{equation}
\begin{aligned}
\hat S_q^\alpha &=
\hat O_{q,c}^\alpha-i\hat O_{q,s}^\alpha,
\\
\hat O_{q,c}^\alpha
&=
\sum_{j=1}^N \cos(2\pi q r_j)\hat S_j^\alpha,
\\
 \hat  O_{q,s}^\alpha
&=
 \sum_{j=1}^N \sin(2\pi q r_j)\hat S_j^\alpha.
\end{aligned}
\end{equation}

As both $\hat O_{q,c}^\alpha$ and $\hat O_{q,s}^\alpha$ are Hermitian, they define valid QFI generators associated with dynamical susceptibilities in analogy to Eq.~\eqref{eq:chi_Sqom}, but based on correlators $\left\langle \hat{O}^\alpha_{q, c}(t)\hat{O}^\alpha_{q,c}(0)\right\rangle$ and $\left\langle \hat{O}^\alpha_{q, s}(t)\hat{O}^\alpha_{q,s}(0)\right\rangle$, respectively. The experimentally accessible witness at a generic momentum is their summed QFI density,
\begin{equation}
\begin{aligned}
f_Q^{c+s}(q,T)
&\equiv
f_Q(\rho,\hat O_{q,c}^\alpha)
+f_Q(\rho,\hat O_{q,s}^\alpha)
\\
&=
2\int_0^\infty \mathrm{d}(\hbar\omega)\,
\tanh\left(\frac{\hbar\omega}{2k_BT}\right)
\left(1-e^{-\hbar\omega/k_BT}\right) \times
\left[
S^\alpha(q,\hbar\omega)+S^\alpha(-q,\hbar\omega)
\right]
\\
&=
2\int_{-\infty}^{\infty}\mathrm{d}(\hbar\omega)\,
\tanh^2\left(\frac{\hbar\omega}{2k_BT}\right) \times
\left[
S^\alpha(q,\hbar\omega)+S^\alpha(-q,\hbar\omega)
\right]
\\
&=
4\int_{-\infty}^{\infty}\mathrm{d}(\hbar\omega)\,
\tanh^2\left(\frac{\hbar\omega}{2k_BT}\right)
S^\alpha(q,\hbar\omega) .
\label{eq:QFI_quadratures}
\end{aligned}
\end{equation}
The second equality follows from detailed balance.  If the system is inversion symmetric, so that $S^\alpha(q,\hbar\omega)=S^\alpha(-q,\hbar\omega)$, then Eq.~\eqref{eq:QFI_quadratures} reduces to a form that is effectively equivalent to  Eq.~\eqref{eq:QFI_RIXS}.  At the self-inverse momenta $q\in\frac{1}{2}\mathbb{Z}$, the sine quadrature vanishes and this is the QFI of the single Hermitian operator $\hat S_q^\alpha$. At a generic momentum it should instead be understood as the sum of the QFIs of the cosine and sine standing-wave spin modulations. For spin $1/2$, their local squared spectral widths obey
\begin{equation}
\sum_{j=1}^N
\left[
\cos^2(2\pi q r_j)+\sin^2(2\pi q r_j)
\right]
=N,
\end{equation}
so the normalization of the multipartite-entanglement bound is unchanged:
$f_Q^{c+s}(q,T)>m$ certifies at least $(m+1)$-partite entanglement.

\section{Impact of $V$ on QFI}\label{sec:Vdep}

\begin{figure}[H]
    \centering
    \includegraphics[width=0.5\linewidth]{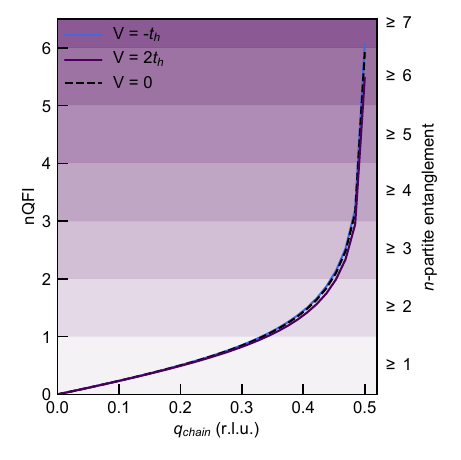}
    \caption{QFI calculated for an EHM chain with an attractive interaction, $V=-t_h$ (solid blue line), a repulsive interaction, $V=2t_h$ (solid purple line), and no nearest-neighbor interaction, $V=0$ (dashed black line).} 
    \label{fig:QFI_Vdep}
\end{figure}

To test the sensitivity of the calculated QFI to the nearest-neighbor interaction $V$, we compare the parameterization used in the main text, $V=-t_h$, with both $V=0$ and the repulsive value $V=2t_h$ used in other parameterizations of one-dimensional cuprates. As shown in Fig.~\ref{fig:QFI_Vdep}, the momentum-dependent QFI changes only weakly across these values of $V$, demonstrating that the inferred entanglement depth is robust to the choice of nearest-neighbor interaction within the parameter range considered here.

\section{Independence of the Projection factor from $U$ and $V$ in RIXS and INS analyses}\label{sec:indepUV}

The dynamical structure factor $S(q, \hbar\omega)$ is defined as:
\begin{equation}
S(q, \hbar\omega) = \frac{1}{\pi N} \, \mathrm{Im} \left\langle G \left| S_{-q}^z \frac{1}{\hat{H} - E_G - \hbar\omega - i\delta} S_q^z \right| G \right\rangle,
\end{equation}
where $|G\rangle$ is the ground state of the Hamiltonian $\hat{H}$ with energy $E_G$, and $S_q^z = \sum_j e^{-i\mathbf{q} \cdot \mathbf{r}_j} S_j^z$ is the spin operator in momentum space. $S(q, \hbar\omega)$ can be rewritten by summing over intermediate states: 
\begin{equation}
S(q, \hbar\omega) = \frac{1}{\pi N} \sum_n \Big| \langle n | S_q^z | G \rangle \Big|^2 \mathrm{Im} \left( \frac{1}{E_n - E_G - \hbar\omega - i\delta} \right),
\end{equation}
where the imaginary part of the resolvent is given by:
\begin{equation}
\mathrm{Im} \left( \frac{1}{E_n - E_G - \hbar\omega - i\delta} \right) = \pi \delta(E_n - E_G - \hbar\omega).
\end{equation}
Integrating over $\hbar\omega$, we find:
\begin{equation}
    \begin{split}
        \int_{-\infty}^\infty & S(q, 
        \hbar\omega) \, d(\hbar\omega)  \\
    & = \frac{1}{N} \sum_n \Big| \langle n | S_q^z | G \rangle \Big|^2 \int_{-\infty}^\infty \delta(E_n - E_G - \hbar\omega) \, d(\hbar\omega) \\
& = \frac{1}{N} \sum_n \Big| \langle n | S_q^z | G \rangle \Big|^2.
    \end{split}
\end{equation}

Similarly, for the \gls*{RIXS} intensity $\mathcal{I}_{\text{RIXS}}(q, \hbar\omega_\mathrm{in}, \hbar\omega)$, we have:
\begin{equation}
    \begin{split}
        \int_{-\infty}^\infty & \mathcal{I}_{\text{RIXS}}(q, \hbar\omega_\mathrm{in}, \hbar\omega) \, d(\hbar\omega) \\ 
        & = \int_{-\infty}^\infty \frac{1}{\pi} \, \mathrm{Im} \Big\langle \Psi_f \Big| \frac{1}{\hat{H} - E_G - \hbar\omega - i\delta} \Big| \Psi_f \Big\rangle \\
        & =  \langle \Psi_f | \Psi_f \rangle.
    \end{split}
\end{equation}

Expanding $\langle \Psi_f | \Psi_f \rangle$, we get:
\begin{equation}
    \begin{split}
        \langle \Psi_f | \Psi_f \rangle & = \sum_{j,j'} e^{i q (r_j - r_{j'})} \cdot\\
        &  \Big\langle G \Big| D_j^\dagger \frac{1}{\hat{H}' - E_G - \hbar\omega_\mathrm{in} + i\Gamma} D_{j'}^\dagger \times \\
        & D_{j'}  \frac{1}{\hat{H}' - E_G - \hbar\omega_\mathrm{in} - i\Gamma}  D_j \Big| G \Big\rangle,
    \end{split}
\end{equation}

and substituting $\hat{H}' - E_G$ as $\hbar\omega$ in the operators
\begin{equation}
\frac{1}{\mathcal{H}' - E_G - \hbar\omega_\mathrm{in} \pm i\Gamma}  = \frac{1}{\hbar\omega - \hbar\omega_\mathrm{in} \pm i\Gamma} \approx \mp \frac{i}{\Gamma},
\end{equation}

we obtain
\begin{align}
\langle \Psi_f | \Psi_f \rangle &\approx \left( \frac{1}{\Gamma} \right)^2 \sum_{j,j'} e^{i q (r_j - r_{j'})} \Big\langle G \Big| D_j^\dagger D_{j'}^\dagger D_{j'} D_j \Big| G \Big\rangle, \\
&\propto \sum_n \Big| \Big\langle n \Big| \sum_j e^{i q r_j} D_j \Big| G \Big\rangle \Big|^2.
\end{align}

The projection factor $\mathcal{N}(q, \Gamma)$ is defined as:
\begin{equation}
\mathcal{N}(q, \Gamma) = \frac{\int_{-\infty}^{\infty} S(q,\hbar\omega)\, d(\hbar\omega)}{\int_{-\infty}^{\infty} \mathcal{I}_{\text{RIXS}}(q,\hbar\omega_{in}, \hbar\omega)\, d(\hbar\omega)}.
\end{equation}

Since both RIXS intensity $\mathcal{I}_{\text{RIXS}}(q,\hbar\omega_{in}, \hbar\omega)$ and $S(q, \hbar\omega)$ depend on similar ground-state correlation functions, the ratio eliminates dependence on Hamiltonian parameters $U$ and $V$. Hence, $\mathcal{N}(q, \Gamma)$ is determined solely by experimental parameters such as core-hole lifetime and matrix elements.

\newpage
\section{Purity of finite-temperature states}\label{sec:purity}

\begin{figure}[h]
    \centering
    \includegraphics[width=0.6\linewidth]{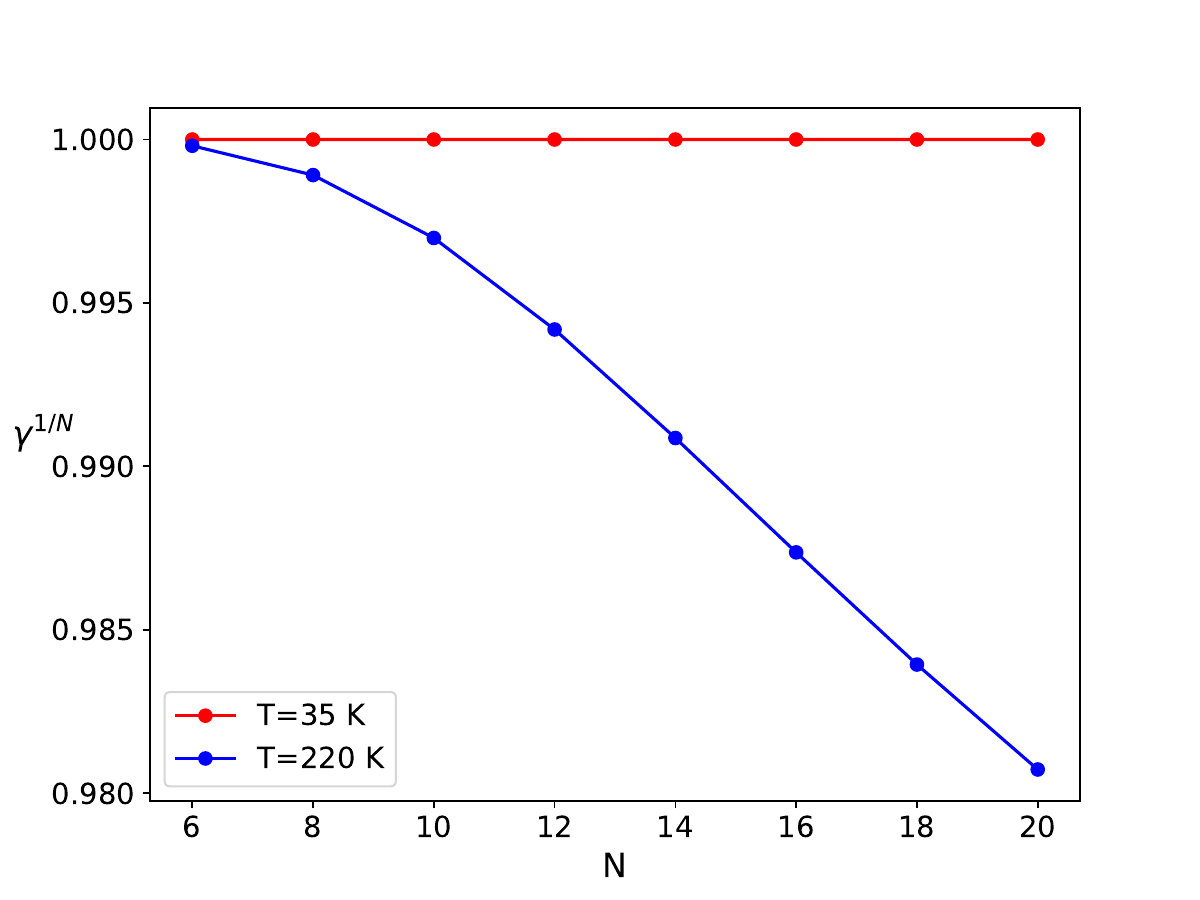}
    \caption{Intensive purity $\mathrm{Tr}(\rho_T^2)^{1/N} \equiv \gamma^{1/N}$ for the Heisenberg model thermal state at temperatures $T = 35 \:\text{K}$ and $T=220\:\mathrm{K}$ for various system sizes $N$ with periodic boundary conditions. Purity is calculated from the eigenvalues of the Heisenberg Hamiltonian using exact diagonalization.}
    \label{fig:purity_ED}
\end{figure}

The multipartite spin entanglement exhibited by the $\text{Sr}_2\text{CuO}_3$ sample at even a relatively high temperature of $220 \:\mathrm{K}$ can be explained by the large superexchange constant $J=250 \:\text{meV}$ of the effective Heisenberg model that describes the material. At $T=220\:\mathrm{K}$, $k_BT/J \approx 0.076 \ll 1$. Therefore, we expect that the thermal (Gibbs) state that describes the state of the system

\begin{equation}\label{eq:gibbs}
    \rho_T = \frac{e^{-\frac{H}{k_BT}}}{Z}
\end{equation}
where $H$ is the Heisenberg Hamiltonian and $Z \equiv \sum_i e^{-E_i/k_BT}$ is the partition function for energies $\{E_i\}$, is not highly mixed and remains close to the ground state at $T=220$ K.

In this section, we present numerical evidence that indicates that the state of the $\text{Sr}_2\text{CuO}_3$ sample is indeed close to being in a pure state at temperatures $35 \ \mathrm{K}$ and $220 \: \mathrm{K}$. A standard measure of how pure a mixed quantum state with density matrix $\rho$ is the \textit{purity}:

\begin{equation}\label{eq:purity}
    \gamma  \equiv \mathrm{Tr}(\rho^2) = \sum_i p_i^2
\end{equation}

where $p_i$ are the eigenvalues of $\rho$. The purity $0 \leq \gamma \leq 1$ measures how mixed a quantum state is, with pure states having $\gamma = 1$ and $\gamma = 0$ for the maximally mixed state. For a thermal state of a gapless Hamiltonian $H$ at fixed temperature $T$ as in Eq. \eqref{eq:gibbs}, the purity is expected to decay exponentially in the system size $N$. This can be seen by observing that the probability $p_i$ of occupying an individual eigenstate decays as $\sim e^{-N}$, while the number of terms in the sum in Eq. \eqref{eq:purity} scales as $2^N$. Consequently, the appropriate intensive quantity that is bounded in $[0,1]$ is $\mathrm{exp}(\mathrm{log}(\gamma)/N) = \gamma^{1/N}$, which we refer to as \textit{intensive purity}. Using exact diagonalization of the Heisenberg Hamiltonian for system sizes up to $N=20$, we compute this quantity  numerically for both temperatures $T = 35 \ \text{K}$ and $T=220 \mathrm{K}$, shown in Fig. \ref{fig:purity_ED}.

The intensive purity at $T=35 \ \mathrm{K}$ can be seen to saturate at $\gamma^{1/N} \approx 1$ as a function of system size $N$ (it is exactly $1$ up to numerical precision). At $T=220 \ \mathrm{K}$, the intensive purity decays with $N$ in the range of accessible system sizes with exact diagonalization techniques. Nonetheless, the intensive purity remains high $\approx 0.98$ at the largest system size $N=20$, although  system sizes that would allow a determination of $\gamma^{1/N}$ as $N\rightarrow \infty$ for $T=220 \ \mathrm{K}$ via extrapolation cannot be accessed numerically at present (the purity is intractable to compute using matrix product state methods for even modest bond dimensions). A numerically computed positive second derivative of the blue curve in Fig. \ref{fig:purity_ED} serves as a weak indication that $\gamma^{1/N}$ indeed asymptotes as $N\rightarrow \infty$ to a non-zero value.

\bibliography{SCO_QFI_RIXS.bib}